\documentclass[aps,pre,superscriptaddress,
               twocolumn,balancelastpage,
               ]{revtex4-1}

\usepackage{xr-hyper}
\usepackage[colorlinks,bookmarks=false,citecolor=blue,linkcolor=blue,urlcolor=blue]{hyperref}

\makeatletter
\newcommand*{\addFileDependency}[1]{
  \typeout{(#1)}
  \@addtofilelist{#1}
  \IfFileExists{#1}{}{\typeout{No file #1.}}
}
\makeatother

\usepackage[all]{hypcap}   

\usepackage{amsmath,amssymb}
\usepackage{graphicx}
\usepackage{enumerate}

\usepackage{verbatim}
\usepackage{color}

\usepackage{placeins}  

\usepackage{braket}
\usepackage{dsfont}

\usepackage{tensor}
\usepackage{tikz}
\usepackage{tikz-network}
\usetikzlibrary{patterns,decorations.pathreplacing}

\definecolor{myred}{RGB}{232,102,102}
\definecolor{myblue}{RGB}{187,187,255}
\definecolor{mygreen}{RGB}{34,139,34}
\definecolor{myorange}{RGB}{255,165,0}
\definecolor{myorange2}{RGB}{240,196,120}
\definecolor{OliveGreen}{RGB}{85,107,47}
\definecolor{NavyBlue}{RGB}{0,0,128}

\renewcommand{\mod}{\text{mod }}

\newcommand{\ui}{\text{i}}

\newcommand{\U}{\mathcal{U}}
\newcommand{\T}{\mathcal{T}}

\newcommand{\N}{\mathcal{N}}
\newcommand{\W}{\mathcal{W}}
\newcommand{\C}{\mathcal{C}}
\newcommand{\R}{\mathcal{R}}
\newcommand{\K}{\mathcal{K}}
\newcommand{\Perm}{\mathds{P}}
\newcommand{\Proj}{\mathcal{P}}

\newcommand{\tr}{\text{tr}}

\newcommand{\sqrtwo}{1.}  
\newcommand{\sqrtwoII}{1.4142135623}

\newcommand{\boundarygate}[2]{
    \draw[thick] (#1 -.5, #2 -0.5) -- (#1 + 0.5, #2 + 0.5);
    \draw[thick] (#1 -0.5, #2 + 0.5) -- (#1 + 0.5, #2 -0.5);
    \draw[thick, fill=myorange, rounded corners=2pt] (#1 -0.4, #2 -0.4) rectangle  (#1 + .4,#2 + .4);
    \draw[thick] (#1 + 0.1,#2 + 0.3) -- (#1 + 0.3,#2 + 0.3);
    \draw[thick] (#1 + 0.3, #2 + 0.1) -- (#1 + 0.3,#2 + 0.3);
}

\newcommand{\transfermatrixgateunrotated}[2]{
    \draw[thick] (#1 -.5, #2 -0.5) -- (#1 + 0.5, #2 + 0.5);
    \draw[thick] (#1 -0.5, #2 + 0.5) -- (#1 + 0.5, #2 -0.5);
    \draw[thick, fill=mygreen, rounded corners=2pt] (#1 -0.4, #2 -0.4) rectangle  (#1 + .4,#2 + .4);
    \draw[thick] (#1 + 0.1,#2 + 0.3) -- (#1 + 0.3,#2 + 0.3);
    \draw[thick] (#1 + 0.3, #2 + 0.1) -- (#1 + 0.3,#2 + 0.3);
}

\newcommand{\Shift}[2]{  
    \draw[thick] (#1 + 0.5, #2 + 0.3) -- (#1 - 0.5, #2 - 0.3);
    \draw[thick] (#1 + 0.5, #2 + 0.3) -- (#1 + 0.5, #2 + 0.5);
    \draw[thick] (#1 - 0.5, #2 - 0.3) -- (#1 - 0.5, #2 - 0.5);
}

\newcommand{\transfermatrixgate}[2]{
    \draw[thick] (#1 -0.5, #2) -- (#1 + 0.5 ,#2);
    \draw[thick] (#1, #2 - 0.5) -- (#1,#2 + 0.5);
    \draw[thick, fill=mygreen, rounded corners=1pt, rotate around={45:(#1, #2)}] (#1 -0.28, #2 -0.28) rectangle  (#1 + .28,#2 + .28);
    \draw[thick] (#1  + 0.15, #2 + 0.12) -- (#1 + 0.27  ,#2);
    \draw[thick] (#1  + 0.15, #2 - 0.12) -- (#1 + 0.27  ,#2);
}

\newcommand{\transfermatrixgateR}[2]{
    \draw[thick] (#1 -0.5, #2) -- (#1 + 0.5 ,#2);
    \draw[thick] (#1, #2 - 0.5) -- (#1,#2 + 0.5);
    \draw[thick, fill=mygreen, rounded corners=1pt, rotate around={45:(#1, #2)}] (#1 -0.28, #2 -0.28) rectangle  (#1 + .28,#2 + .28);
    \draw[thick] (#1  + 0.12, #2 + 0.15) -- (#1  ,#2 + 0.27);
    \draw[thick] (#1  - 0.12, #2 + 0.15) -- (#1, #2 + 0.27);
}

\newcommand{\SwapR}[2]{
    \draw[thick] (#1 -0.5, #2) -- (#1 + 0.5 ,#2);
    \draw[thick] (#1, #2 - 0.5) -- (#1,#2 + 0.5);
}

\newcommand{\Swap}[2]{
    \draw[thick] (#1 -0.5, #2 - 0.5) -- (#1 + 0.5 ,#2 + 0.5);
    \draw[thick] (#1 + 0.5, #2 - 0.5) -- (#1 - 0.5, #2 + 0.5);
}

\newcommand{\Identity}[2]{
    \draw[thick] (#1, #2 - 0.5) -- (#1, #2 + 0.5);
}

\newcommand{\IdentityII}[2]{
\draw[thick] (#1, #2 - 0.5/\sqrtwo) -- (#1, #2 + 0.5/\sqrtwo);
}

\newcommand{\IdState}[2]{  
    \draw[thick, fill=white] (#1,#2) circle (0.12cm);
}

\newcommand{\IdStateII}[2]{  
    \draw[thick, fill=white] (#1,#2) circle (\sqrtwoII*0.12cm);
}

\newcommand{\localOperator}[2]{  
    \draw[thick, fill=black] (#1,#2) circle (0.12cm);
}

\newcommand{\localOperatorII}[2]{  
    \draw[thick, fill=black] (#1,#2) circle (\sqrtwoII*0.12cm);
}

\newcommand{\initialOperator}[2]{
    \draw[thick] (#1, #2) -- (#1 + 0.5, #2 + 0.3);
    \draw[thick] (#1 + 0.5, #2 + 0.3) -- (#1 + 0.5, #2 + 0.5);
    \draw[thick] (#1 - 0.2, #2) -- (#1, #2);
    \localOperator{#1 - 0.2}{#2}
}

\newcommand{\initialOperatorId}[2]{
    \draw[thick] (#1, #2) -- (#1 + 0.5, #2 + 0.3);
    \draw[thick] (#1 + 0.5, #2 + 0.3) -- (#1 + 0.5, #2 + 0.5);
    \draw[thick] (#1 - 0.2, #2) -- (#1, #2);
    \IdState{#1 - 0.2}{#2}
}

\newcommand{\finalOperator}[2]{
    \draw[thick] (#1, #2) -- (#1 - 0.5, #2 - 0.3);
    \draw[thick] (#1 - 0.5, #2 - 0.3) -- (#1 - 0.5, #2 - 0.5);
    \draw[thick] (#1 + 0.2, #2) -- (#1, #2);
    \localOperator{#1 + 0.2}{#2}
}
\newcommand{\finalOperatorId}[2]{
    \draw[thick] (#1, #2) -- (#1 - 0.5, #2 - 0.3);
    \draw[thick] (#1 - 0.5, #2 - 0.3) -- (#1 - 0.5, #2 - 0.5);
    \draw[thick] (#1 + 0.2, #2) -- (#1, #2);
    \IdState{#1 + 0.2}{#2}
}

\newcommand{\HIDDEN}[1]{}

\makeatletter
\let\Hy@backout\@gobble
\makeatother

\begin{document}

\title{Boundary Chaos}

\author{Felix Fritzsch}
\affiliation{Physics Department, Faculty of Mathematics and Physics,
    University of Ljubljana, Ljubljana, Slovenia}

\author{Toma\v{z} Prosen}
\affiliation{Physics Department, Faculty of Mathematics and Physics,
    University of Ljubljana, Ljubljana, Slovenia}

\date{\today}

\begin{abstract}
    Scrambling in many-body quantum systems causes initially local observables to spread uniformly
    over the whole available Hilbert space  under unitary dynamics, which in lattice systems causes exponential suppression of dynamical correlation functions with system size.
    Here, we present a perturbed free quantum circuit model, in which ergodicity is induced by
    an impurity interaction placed on the system's boundary, that allows for demonstrating the underlying 
    mechanism.
    This is achieved by mapping dynamical correlation functions of local operators acting at the boundary to a partition 
    function with complex weights defined on a two dimensional lattice with a helical topology.
    We evaluate this partition function in terms of transfer matrices,
    which allow for numerically treating system sizes far beyond what is accessible by exact 
    diagonalization
    and whose spectral properties determine the asymptotic scaling of correlations.
    Combining analytical arguments with numerical results we show that for impurities which remain 
    unitary under partial transpose correlations are exponentially suppressed with system size in a
    particular scaling limit.
    In contrast for generic impurities or generic locations of the local operators correlations show persistent revivals with a period given by the system size.
\end{abstract}

\maketitle

Spatiotemporal correlation functions provide the key diagnostic tool 
for studying spatially extended
complex quantum many-body systems, both theoretically \cite{AltSim2010,Set2006} and experimentally \cite{BloDalZwe2008,CheBarPol2012,RicGonLee2014}, and most recently also in quantum simulations \cite{CheBohFraGaeGreHanLeeTobHayNeyStuPotFos2021:p}.
They give rise to a notion of ergodicity in quantum systems 
\cite{BerKosPro2019,AraRatLak2021}
and allow for quantifying scrambling of local quantum information \cite{HayPre2007,SekSus2008,MalSheSta2016,Swi2018}.
Moreover, two-point correlation functions determine transport coefficient of conserved currents \cite{Mah2010} as well as the relaxation dynamics of local operators and their approach to thermal equilibrium \cite{AleKafPolRig2016}.
More precisely, the relaxation dynamics of auto-correlation functions encodes statistical properties of the operators' matrix elements
in the energy eigenbasis and hence yields a convenient tool to study the validity of the famous eigenstate thermalization hypothesis (ETH) \cite{Deu1991,Sre1994,AleKafPolRig2016}.
The latter conjectures universal statistics of these matrix elements.
In order for ETH to hold true correlation functions of large finite systems are required to relax to a value which is exponentially suppressed with the system size.

The computation of spatiotemporal correlation functions
 is a challenging task in general due to entangling dynamics of generic interacting many-body quantum systems, which limits the applicability of numerical methods to relatively small systems.
In contrast, recent progress in identifying exactly solvable chaotic many-body systems provided numerous examples where correlation functions can be computed exactly for all times in the thermodynamic limit or for extensive finite times in finite systems.
Most notably in quantum circuit models~\cite{BerKosPro2019,GopLam2019,ClaLam2021,GutBraAkiWalGuh2020,KosBerPro2021,ClaHerLam2021:p,AraRatLak2021}
which exhibit a duality between space and time, so-called dual unitarity \cite{AkiWalGutGuh2016,BerKosPro2019},
correlation functions are determined by single-site quantum channels as the rest of the system plays the role of a perfect markovian bath.
These systems can be robust upon small perturbations which break space-time duality \cite{KosBerPro2021} and hence give rise to an ideal testbed for studying properties of ergodic many-body systems.

Here we present a setup which can be thought of as a toy model for a non-interacting (aka free) system in which chaos and ergodicity, i.e. decay of correlations with time, are induced by a local impurity.
Chaos and ergodicity have been observed in interacting integrable systems upon an integrability breaking local impurity \cite{San2004,BarPreMetZot2009,TorSan2014,BreMasRigGoo2018,BreLeBGooRig2020,PanClaCamPolSel2020,Znidaric} but it is less clear, whether this is also the case if the unperturbed system is non-interacting. 
Conceptually, studying perturbed free systems might allow for exactly integrating out the trivial free dynamics in analogy to the concept of Poincar\'e sections used in classical billiard's dynamics~\cite{Ott2002} and the associated quantized Poincar\'e transfer operator~\cite{Bog1992,Pro1995}, which reduces the effective dimension of phase space via often trivial integration between subsequent sections (boundary collisions in billiards).
In the context of many-body systems the integrated free dynamics is in close analogy with the influence matrix approach \cite{LerSonAba2021} as it describes the influence of the trivial bulk dynamics on the nontrivial part of the systems exactly even for finite systems.

In this paper we apply this strategy to generally chaotic Floquet quantum circuits in which two-point correlation functions between local operators can be computed without approximations for times up to a fixed number of multiples of system size for arbitrarily large systems. We can thus target a niche regime between the realm of exact diagonalization in finite systems and exact results in the thermodynamic limit. 
Specifically, we consider a free quantum circuit composed as a brickwork of swap gates in
which ergodicity and scrambling is induced by perturbing the circuit with an impurity interaction placed on the system's boundary.
We map the correlation functions of local operators at the boundary subject to the Floquet dynamics at time $t$ induced by the circuit of size $L$ to a partition function defined on a quasi-one-dimensional $\sim t/L \times L$ lattice with a helical topology.
The latter is evaluated in terms of transfer matrices which for the considered time regime are of much smaller dimensionality than the actual Floquet operator.
We derive the asymptotic behavior of correlations for large $L$ and fixed $t/L$
from the leading parts of the transfer matrices' spectra and find exponential suppression of correlations in $L$ for impurity interactions that remain unitary under partial transposition.
Supported by the numerical computation of the leading eigenvalues for increasing $t/L$ we conjecture correlations to be exponentially suppressed for all times $t>L$ in this case.
In contrast, for generic impurities or generic locations of the local operators we find persistent revivals of correlations with period $L$, i.e. around integer $t/L$.

\section{Boundary chaos}

The models we consider are built as free brickwork quantum circuits composed of swap gates on a lattice of size $L + 1$ with open boundary conditions.
To each lattice site $x \in \{0, 1, \ldots, L\}$ we assign a local Hilbert
space $\mathds{C}^q$ of dimension $q$ giving rise to a total Hilbert space
$\bigotimes_x \mathds{C}^q \cong \mathds{C}^N$, where $N=q^{L+1}$.
We argue that ergodicity and scrambling may be induced by replacing the swap gate acting on the first two lattice sites by an impurity interaction, i.e., a
generic unitary $U \in \text{U}(q^2)$.
We refer to this setup as {\em boundary chaos}.
Formally, we define the Floquet operator corresponding to this circuit layout as $\U=\U_2\U_1 \in \text{End}\left(\mathds{C}^{N}\right)$
with
\begin{align}
\U_1 = \prod_{i=1}^{\lfloor L/2 \rfloor} P_{2i-1, 2i}\,, \quad  
\U_2 = U_{0, 1}\prod_{i=1}^{\lfloor (L-1)/2 \rfloor}P_{2i, 2i+1}\,,
\label{eq:circuit_def}
\end{align}
where $P_{i,j}$ ($U_{i,j}$) denotes
unitary gate acting nontrivially as the swap $P$ (interaction $U$) on sites $i$,$j$, and trivially otherwise.
The resulting circuit is found to exhibit quantum chaos in the spectral sense as its spectral 
statistics match random matrix theory for typical $U$, see App.~\ref{S-II}.
Here, however, we are interested in ergodicity and mixing in the sense of the decay of dynamical correlations
\begin{align}
C_{ab}(t) = \frac{1}{N}\tr\left(\U^{-t}a_0\U^t b_0 \right)
\label{eq:correlation}
\end{align}
between local operators $a_0$ and $b_0$ acting 
as traceless Hermitian operators $a$ and $b \in\text{End}(\mathds{C}^q)$
on lattice site $0$, respectively, and trivially otherwise.
We treat local operators acting nontrivial on arbitrary lattice sites in Sec.~\ref{sec:generic_locations}.
Note, that the normalized trace corresponds to taking averages with respect to the invariant infinite
temperature state, which constitutes the natural, and in generic case of ergodicity, unique equilibrium state in Floquet systems.

We cast the Heisenberg time evolution of operators into a quantum circuit formulation with enlarged local Hilbert spaces
allowing for a diagrammatic representation: The so called folded picture 
\cite{BanHasVerCir2009,MueCirBan2012}
which we introduce in the following.
To this end we employ the unitary (w.r.t.~ the Hilbert-Schmidt scalar product on $\text{End}(\mathds{C}^q)$) 
operator-to-state mapping by bilinear extension of
$\text{End}(\mathds{C}^q) \ni \ket{m}\!\bra{n} \mapsto \ket{m}\!\otimes\!\ket{n} 
\in \mathds{C}^{q}\otimes \mathds{C}^q \cong \mathds{C}^{q^2}$. Here $\{\ket{n}|n = 1, \ldots, q \}$ is the canonical basis of $\mathds{C}^q$.
We write $\ket{a}$
for the image of an operator $a$ and, specifically, $\ket{\circ}$ for the image of $q^{-1/2}\mathds{1}_{q}$. 
Via tensor multiplication the vectorization mapping extends to an isomorphism $\text{End}( (\mathds{C}^q)^{\otimes (L+1)}) \cong
(\mathds{C}^{q^2})^{\otimes(L+1)}$ while
the Heisenberg time evolution is cast into a
quantum circuit formulation.
This yields a Floquet operator $\W=\W_1\W_2$ acting on the lattice
(of size $L+1$) of $q^2$-dits, with
\begin{align}
\W_2 = W_{0,1}\prod_{i}S_{2i, 2i+1}  \,, \quad
\W_1 = \prod_{i} S_{2i-1, 2i}\,,
\end{align}
where the folded local gates are defined as 
$S = P \otimes P$
and
$W = U^\dagger \otimes U^{\text{T}}$.

We obtain a diagrammatic representation \cite{Pen1971} for correlation functions~\eqref{eq:correlation} by introducing the graphical notation
$ 
S = 
\begin{tikzpicture}[baseline=-3., scale=0.4]
\Swap{0.}{0.}
\end{tikzpicture}
$ 
and
$W= 
\begin{tikzpicture}[baseline=-3., scale=0.4]
\boundarygate{0.}{0.}
\end{tikzpicture}\,,
$ 
for the local gates of the circuit as well as 
$\ket{a} = 
\begin{tikzpicture}[baseline=-1.5, scale=0.425]
 \draw[thick] (0., 0.) -- (0., 0.5);
\localOperatorII{0.}{0.}
\Text[x=0.5,y=-0.1]{$a$}
\end{tikzpicture}$
and
$ 
\ket{\circ} = 
\begin{tikzpicture}[baseline=-1.5, scale=0.425]
\draw[thick] (0., 0.) -- (0., 0.5);
\IdStateII{0.}{0.}
\end{tikzpicture}\,$,
for vectorized operators. 
Then the circuit formulation of Heisenberg time evolution is used to recast Eq.~\eqref{eq:correlation} as
a particular tensor network with nontrivial boundary 
\begin{align}
C_{ab}(t) = \bra{b_0}\W^t\ket{a_0} =
\begin{tikzpicture}[baseline=25, scale=0.425]
\foreach \j in {0, 2, 4}
{
    \boundarygate{0./\sqrtwo}{\j/\sqrtwo}
}
\foreach \j in {0, 2, 4}
\foreach \i in {1}
{
    \Swap{2*\i/\sqrtwo}{\j/\sqrtwo}
}
\foreach \j in {1, 5}
\foreach \i in {1, 2.6}
{
    \Swap{2*\i/\sqrtwo - 1/\sqrtwo}{\j/\sqrtwo}
}
\foreach \j in {1, 5}
\foreach \i in {0}
{
    \IdentityII{\i/\sqrtwo - 0.5/\sqrtwo}{\j/\sqrtwo}
}
\foreach \j in {0, 2, 4}
\foreach \i in {5.2}
{
    \IdentityII{\i/\sqrtwo - 0.5/\sqrtwo}{\j/\sqrtwo}
}
\foreach \i in {0.5, 4.5}
{
\draw[thick] (2.5/\sqrtwo, \i/\sqrtwo) -- (2.5/\sqrtwo + 0.2, (\i/\sqrtwo + 0.2);
}
\foreach \i in {1.5, 5.5}
{
\draw[thick] (2.5/\sqrtwo, \i/\sqrtwo) -- (2.5/\sqrtwo + 0.2, (\i/\sqrtwo - 0.2);
}
\foreach \i in {0.5, 2.5, 4.5}
{
\draw[thick] (3.7/\sqrtwo, \i/\sqrtwo) -- (3.7/\sqrtwo - 0.2, (\i/\sqrtwo - 0.2);
}
\foreach \i in {-0.5, 1.5, 3.5}
{
\draw[thick] (3.7/\sqrtwo, \i/\sqrtwo) -- (3.7/\sqrtwo - 0.2, (\i/\sqrtwo + 0.2);
}
\node at (-0.5/\sqrtwo, 3./\sqrtwo)[circle,fill,inner sep=0.4pt]{};
\node at (-0.5/\sqrtwo, 3./\sqrtwo - 0.15)[circle,fill,inner sep=0.4pt]{};
\node at (-0.5/\sqrtwo, 3./\sqrtwo + 0.15)[circle,fill,inner sep=0.4pt]{};
\node at (0.5/\sqrtwo, 3./\sqrtwo)[circle,fill,inner sep=0.4pt]{};
\node at (0.5/\sqrtwo, 3./\sqrtwo - 0.15)[circle,fill,inner sep=0.4pt]{};
\node at (0.5/\sqrtwo, 3./\sqrtwo + 0.15)[circle,fill,inner sep=0.4pt]{};
\node at (1.5/\sqrtwo, 3./\sqrtwo)[circle,fill,inner sep=0.4pt]{};
\node at (1.5/\sqrtwo, 3./\sqrtwo - 0.15)[circle,fill,inner sep=0.4pt]{};
\node at (1.5/\sqrtwo, 3./\sqrtwo + 0.15)[circle,fill,inner sep=0.4pt]{};
\node at (4.7/\sqrtwo, 3./\sqrtwo)[circle,fill,inner sep=0.4pt]{};
\node at (4.7/\sqrtwo, 3./\sqrtwo - 0.15)[circle,fill,inner sep=0.4pt]{};
\node at (4.7/\sqrtwo, 3./\sqrtwo + 0.15)[circle,fill,inner sep=0.4pt]{};
\node at (3.7/\sqrtwo, 3./\sqrtwo)[circle,fill,inner sep=0.4pt]{};
\node at (3.7/\sqrtwo, 3./\sqrtwo - 0.15)[circle,fill,inner sep=0.4pt]{};
\node at (3.7/\sqrtwo, 3./\sqrtwo + 0.15)[circle,fill,inner sep=0.4pt]{};
\node at (2.5/\sqrtwo, 3./\sqrtwo)[circle,fill,inner sep=0.4pt]{};
\node at (2.5/\sqrtwo, 3./\sqrtwo - 0.15)[circle,fill,inner sep=0.4pt]{};
\node at (2.5/\sqrtwo, 3./\sqrtwo + 0.15)[circle,fill,inner sep=0.4pt]{};
\node at (3.1/\sqrtwo, 5./\sqrtwo)[circle,fill,inner sep=0.4pt]{};
\node at (3.1/\sqrtwo - 0.15, 5./\sqrtwo)[circle,fill,inner sep=0.4pt]{};
\node at (3.1/\sqrtwo + 0.15, 5./\sqrtwo)[circle,fill,inner sep=0.4pt]{};
\node at (3.1/\sqrtwo, 4./\sqrtwo)[circle,fill,inner sep=0.4pt]{};
\node at (3.1/\sqrtwo - 0.15, 4./\sqrtwo)[circle,fill,inner sep=0.4pt]{};
\node at (3.1/\sqrtwo + 0.15, 4./\sqrtwo)[circle,fill,inner sep=0.4pt]{};
\node at (3.1/\sqrtwo, 2./\sqrtwo)[circle,fill,inner sep=0.4pt]{};
\node at (3.1/\sqrtwo - 0.15, 2./\sqrtwo)[circle,fill,inner sep=0.4pt]{};
\node at (3.1/\sqrtwo + 0.15, 2./\sqrtwo)[circle,fill,inner sep=0.4pt]{};
\node at (3.1/\sqrtwo, 1./\sqrtwo)[circle,fill,inner sep=0.4pt]{};
\node at (3.1/\sqrtwo - 0.15, 1./\sqrtwo)[circle,fill,inner sep=0.4pt]{};
\node at (3.1/\sqrtwo + 0.15, 1./\sqrtwo)[circle,fill,inner sep=0.4pt]{};
\node at (3.1/\sqrtwo, 0./\sqrtwo)[circle,fill,inner sep=0.4pt]{};
\node at (3.1/\sqrtwo - 0.15, 0./\sqrtwo)[circle,fill,inner sep=0.4pt]{};
\node at (3.1/\sqrtwo + 0.15, 0./\sqrtwo)[circle,fill,inner sep=0.4pt]{};
%
\foreach \j in {-0.8}
\foreach \i in {1,2, 3, 4.2, 5.2}
{
    \draw[thick] (\i/\sqrtwo - 0.5/\sqrtwo , \j/\sqrtwo +  0.3/\sqrtwo) -- (\i/\sqrtwo - 0.5/\sqrtwo, \j/\sqrtwo);
    \IdStateII{\i/\sqrtwo -0.5/\sqrtwo}{\j/\sqrtwo }
}
\foreach \j in {5.8}
\foreach \i in {1,2, 3, 4.2, 5.2}
{
    \draw[thick] (\i/\sqrtwo - 0.5/\sqrtwo , \j/\sqrtwo -  0.3/\sqrtwo) -- (\i/\sqrtwo - 0.5/\sqrtwo, \j/\sqrtwo);
    \IdStateII{\i/\sqrtwo -0.5/\sqrtwo}{\j /\sqrtwo}
}
\draw[thick] ( - 0.5/\sqrtwo , -0.8/\sqrtwo) -- ( - 0.5/\sqrtwo, -0.5/\sqrtwo);
\localOperatorII{ -0.5/\sqrtwo}{-0.8/\sqrtwo}
\Text[x=-0.9/\sqrtwo,y=-1./\sqrtwo]{$a$}
\draw[thick] ( - 0.5/\sqrtwo , 5.8/\sqrtwo -  0.3/\sqrtwo) -- ( - 0.5/\sqrtwo, 5.8/\sqrtwo);
\localOperatorII{ -0.5/\sqrtwo}{5.8/\sqrtwo}
\Text[x=-0.9/\sqrtwo,y=6./\sqrtwo]{$b$} 
\draw [thick, stealth-stealth](5.3/\sqrtwo, 5.6/\sqrtwo) -- (5.3/\sqrtwo, -0.7/\sqrtwo);
\Text[x=5.6/\sqrtwo ,y=2.5/\sqrtwo]{\footnotesize $t$}
\draw [thick, stealth-stealth](-0.4, -1.2) -- (4.7, -1.2);
\Text[x=2.2,y=-1.6]{\footnotesize $L + 1$}
\end{tikzpicture}.
\label{eq:correlations_folded}
\end{align}

\section{Mapping to helical circuit}
In order to evaluate Eq.~\eqref{eq:correlations_folded} we map the $(L+1)\times t$ tensor network representing the correlation function to a partition function on a $(\tau + 1) \times L$ lattice with the topology
of a helix.
The non-negative integer $\tau$ is defined by writing time 
as $t = L \tau + \delta$ with remainder $\delta \in \{0, 1, \ldots L-1 \}$.
The partition function can then be expressed by
transfer matrices $\T_\tau \in
\text{End}((\mathds{C}^{q^2})^{ \otimes \tau})$ given as
matrix product operators whereas the 
helix topology as well as the initial and final operators are encoded by a shift operator
$\C_{ab, \tau} \in \text{End}((\mathds{C}^{q^2})^{ \otimes \tau})$ with fixed boundary conditions $a$ and $b$, respectively.
Both $\T_\tau$ and $\C_{ab, \tau}$ are most conveniently defined using their diagrammatic 
representation (here for $\tau=4$) as
\begin{align}
\T_\tau & =  \, \, 
\begin{tikzpicture}[baseline=-1.5, scale=0.55]
\foreach \x in {0,...,3}
\foreach \y in {0}
{
    \transfermatrixgate{\x}{\y}
}
\foreach \y in {0}
{
    \draw[thick] (-0.5, \y) -- (- 0.7, \y);
    \draw[thick] (3.5, \y) -- (3.7, \y);    
    \IdState{-0.7}{\y}
    \IdState{3.7}{\y}
}
\end{tikzpicture}
\text{, where} \quad
\begin{tikzpicture}[baseline=-1.5, scale=0.389]
\transfermatrixgateunrotated{0}{0}
\end{tikzpicture}
=\begin{tikzpicture}[baseline=-1.5, scale=0.389]
\boundarygate{0}{0.5}
\Swap{0}{-0.5}
\end{tikzpicture} = V, \label{eq:transfer_matrix} \\
\C_{ab, \tau} & =
\begin{tikzpicture}[baseline=-1.5, scale=0.55]
\foreach \x in {0,...,2}
{
    \Shift{\x+0.5}{0.}
}
\initialOperator{-0.5}{0.}
\finalOperator{3.5}{0.}
\Text[x=-0.8,y=-0.4]{$a$}
\Text[x=3.8,y=-0.4]{$b$}
\draw [thick, stealth-stealth](-0., -0.7) -- (3., -0.7);
\Text[x=1.5,y=-0.95]{\footnotesize $\tau$}
\end{tikzpicture}.  \label{eq:shift_operator}
\end{align}
Using this definitions we obtain for the dynamical correlation function~\eqref{eq:correlation}:
\begin{align}
C_{ab}(t) & = \tr\left(\left[\T_\tau^{L-\delta} \otimes \mathds{1}_{q^2}
\right]\T_{\tau+1}^\delta \C_{ab, \tau + 1}\right) 
\label{eq:correlations_via_transfer_matrix} \\
& =
\begin{tikzpicture}[baseline=22, scale=0.6]
 \foreach \x in {0,...,3}
{
    \draw[thick] (\x, -0.5 ) arc (180:360: 0.15);
    \draw[thick] (\x, 3.5 ) arc (360:180: -0.15);
    \draw[thick, color=gray, dashed] (\x + 0.3, -0.5) -- (\x + 0.3, 3.5);    
}
\foreach \x in {0,...,2}
{
    \Shift{\x+0.5}{0.}
}
\foreach \x in {0,...,3}
\foreach \y in {1,2}
{
    \transfermatrixgate{\x}{\y}
}
\foreach \y in {1,2}
{
    \draw[thick] (-0.5, \y) -- (- 0.7, \y);
    \draw[thick] (3.5, \y) -- (3.7, \y);    
    \IdState{-0.7}{\y}
    \IdState{3.7}{\y}
}
\foreach \x in {0,...,2}
\foreach \y in {3}
{
    \transfermatrixgate{\x}{\y}
}
\Identity{3.}{3.}
\foreach \y in {3}
{
    \draw[thick] (-0.5, \y) -- (- 0.7, \y);
    \draw[thick] (2.5, \y) -- (2.7, \y);    
    \IdState{-0.7}{\y}
    \IdState{2.7}{\y}
}
\initialOperator{-0.5}{0.}
\finalOperator{3.5}{0.}
\Text[x=-0.8,y=-0.4]{$a$}
\Text[x=3.8,y=-0.4]{$b$}
\draw [thick, stealth-stealth](4.2, 2.15) -- (4.2, 0.85);
\Text[x=4.45,y=1.5]{\footnotesize$\delta$}
\draw [thick, stealth-stealth](4.9, 3.15) -- (4.9, 0.85);
\Text[x=5.15,y=2.]{\footnotesize$L$}
\draw [thick, stealth-stealth](-0.2, -0.8) -- (3.2, -0.8);
\Text[x=1.6,y=-1.05]{\footnotesize$\tau + 1$}
\end{tikzpicture}, \label{eq:correlations_via_transfer_matrix_diagramm} 
\end{align}
which can be verified by tracing the wires corresponding to the swap gates
in the diagrammatic representation~\eqref{eq:correlations_folded}; see App.~\ref{S-III} for a formal derivation.
Intuitively, the two different tensor network representations~\eqref{eq:correlations_folded}~and~\eqref{eq:correlations_via_transfer_matrix_diagramm} are related as follows:
The nontrivial local operator $a$ in the bottom left corner of the network~\eqref{eq:correlations_folded} might be scattered into the bulk (swap) part of the network by the impurity at $t=1$.
Subsequently it travels freely forth and back through the system in time $t=L$ until the corresponding wire runs into an impurity interaction at the boundary again.
In the helical network~\eqref{eq:correlations_via_transfer_matrix_diagramm}
this process corresponds to the operator travelling from left to right.
Consequently, the transfer matrices $\T_\tau$ describe the process of local operators freely traveling forth and back through the bulk of the network~\eqref{eq:correlations_folded} $\tau$ times along the wires corresponding to the swap gates and being scattered back, whenever these wires hit the impurity at the boundary.
Additionally, instead of being scattered into the bulk, the local operator $a$ might just travel along the boundary in the network~\eqref{eq:correlations_folded}.
This corresponds to the operator travelling from bottom to top in the helical network.

From the computational complexity point of view, we replaced direct computation of correlation functions, which is linear in $t$ and
exponential in $L$, by transfer matrix contraction of 
the tensor network~\eqref{eq:correlations_via_transfer_matrix_diagramm}, which is linear in $L$ and exponential in $\tau\approx t/L$.
Hence
Eqs.~\eqref{eq:correlations_via_transfer_matrix}~and~\eqref{eq:correlations_via_transfer_matrix_diagramm}
allow to efficiently determine the initial dynamics
of correlation functions up to times $t=\tau L$ for not too large fixed $\tau$ for system sizes 
$L$ much larger than what is accessible by direct methods.
Figure~\ref{fig:correlations} depicts a representative example for $L=200$.

Moreover, the above 
Eqs.~\eqref{eq:correlations_via_transfer_matrix}~and~\eqref{eq:correlations_via_transfer_matrix_diagramm}
suggest that 
the asymptotic scaling of $C_{ab}(\tau L + \delta)$
for both $L - \delta$ and $\delta$ being large is dominated by the leading eigenvalues 
of $\T_{\tau}$ and $\T_{\tau + 1}$.
Hence we describe the spectral properties of the transfer matrices
$\T_\tau$ in the following.
$\T_\tau$ is a vectorization of a quantum channel, a non-expanding map,
with spectrum $\text{spec}(\T_\tau)$ contained in the complex unit disk~\cite{KosBerPro2021}.
Its eigenvalues are either real or come in complex conjugate pairs, and
$\T_\tau$ is in general not diagonalizable, but exhibits nontrivial Jordan blocks.
Unitarity of interaction $U$ implying unitality of the folded gate $V$, i.e.,
\begin{align}
\begin{tikzpicture}[baseline=-3, scale=0.425]
\transfermatrixgateunrotated{0}{0}
\IdStateII{-0.6}{-0.6}
\IdStateII{0.6}{-0.6}
\end{tikzpicture}
=
\begin{tikzpicture}[baseline=-3, scale=0.425]
\draw[thick] (-0.27, -0.27) -- (-0.27, 0.27);
\draw[thick] (0.27, -0.27) -- (0.27, 0.27);
\IdStateII{-0.27}{-0.27}
\IdStateII{0.27}{-0.27}
\end{tikzpicture}
\quad \text{and} \quad
\begin{tikzpicture}[baseline=-3, scale=0.425]
\transfermatrixgateunrotated{0}{0}
\IdStateII{-0.6}{0.6}
\IdStateII{0.6}{0.6}
\end{tikzpicture}
=
\begin{tikzpicture}[baseline=-3, scale=0.425]
\draw[thick] (-0.27, -0.27) -- (-0.27, 0.27);
\draw[thick] (0.27, -0.27) -- (0.27, 0.27);
\IdStateII{-0.27}{0.27}
\IdStateII{0.27}{0.27}
\end{tikzpicture}
\label{eq:unitality}
\end{align}
guarantees that
there is always the trivial (left and right) eigenvector
$\ket{\circ}^{\otimes \tau}$ with trivial eigenvalue $1 \in \text{spec}(\T_\tau)$.
Moreover, unitality of $V$ implies that 
the spectra of transfer matrices grow with
$\tau$, i.e., $\text{spec}\left(\T_\tau\right) \subseteq \text{spec}\left(\T_{\tau+1}\right)$.

\section{T-dual impurities}
In order to be able to analyze the nontrivial eigenvectors as well we first assume folded gates
$V$ to be dual-unitary~\cite{BerKosPro2019}.
More precisely, upon exchanging the role of space and time the folded gate $V$ remains unitary (unital), which might be expressed as
\begin{align}
\begin{tikzpicture}[baseline=-3, scale=0.425]
\transfermatrixgateunrotated{0}{0}
\IdStateII{-0.6}{-0.6}
\IdStateII{-0.6}{0.6}
\end{tikzpicture}
=
\begin{tikzpicture}[baseline=-3, scale=0.425]
\draw[thick] (-0.27, -0.27) -- (0.27, -0.27);
\draw[thick] (-0.27, 0.27) -- (0.27, 0.27);
\IdStateII{-0.27}{0.27}
\IdStateII{-0.27}{-0.27}
\end{tikzpicture}
\quad \text{and} \quad
\begin{tikzpicture}[baseline=-3, scale=0.425]
\transfermatrixgateunrotated{0}{0}
\IdStateII{0.6}{-0.6}
\IdStateII{0.6}{0.6}
\end{tikzpicture}
=
\begin{tikzpicture}[baseline=-3, scale=0.425]
\draw[thick] (-0.27, -0.27) -- (0.27, -0.27);
\draw[thick] (-0.27, 0.27) -- (0.27, 0.27);
\IdStateII{0.27}{-0.27}
\IdStateII{0.27}{0.27}
\end{tikzpicture}.
\label{eq:dual_unitality}
\end{align}
Note, that dual unitarity of $V$ is equivalent to the impurity interaction $U$ being T-dual \cite{AraRatLak2021}, i.e., the
partial transpose with respect to the first (or equivalently the second) site of $U$ being unitary.
Such gates can be parameterized as \cite{BerKosPro2019}
\begin{align}
U= \left(u_+ \otimes u_-\right) \exp\left(\ui J \sigma_{q^2-1} \otimes \sigma_{q^2-1} \right) \left(v_+ \otimes v_- \right).
\label{supp_eq:parametrization_T_dual}
\end{align}
with $\sigma_i$ the generalized Gell-Mann matrices, $J \in \left[0, \pi/4\right]$ and
$u_\pm, v_\pm \in \text{U}(q)$.
This parameterization is exhaustive for $q=2$ only.

For T-dual impurities and hence dual-unitary gates $V$ we observe that $\T_\tau$ is generically diagonalizable.
Moreover, the structure of nontrivial eigenvectors of $\T_\tau$ can be described in some detail.
For the right (left) eigenvector
$\ket{r_\lambda}$ ($\bra{l_\lambda}$) with eigenvalue $\lambda$,
$\bra{l_\lambda}\T_\tau = \lambda\bra{l_\lambda}$,
$\T_\tau\ket{r_\lambda} = \lambda\ket{r_\lambda}$,
also the vector
\begin{align}
   \ket{r_\lambda, s} = \ket{\circ}^{\otimes s}\otimes
   \ket{r_\lambda} \otimes \ket{\circ}^{\otimes \rho - \tau - s}
   \label{eq:eigenvectors}
\end{align}
(and analogous expression for $\bra{l_\lambda,s}$),
with $s \in \{0, \ldots, \rho-\tau\}$ is a right (left) eigenvector of 
$\T_{\rho}$ for $\rho > \tau $ corresponding to the same eigenvalue.
Consequently, $\text{spec}(\T_\tau) \subseteq \text{spec}(\T_{\tau+1})$.
For each
eigenvalue $\lambda$ there is thus $\tau_\lambda$ such that
$\lambda \in \text{spec}(\T_{\tau_\lambda})$ but $\lambda \notin \text{spec}(\T_{\tau_\lambda -1})$.
The corresponding eigenvector (eigenoperator) has full support on the lattice on which $\T_{\tau_\lambda}$ acts.
We use the notation $\ket{r_\lambda}$ ($\bra{l_\lambda}$) exclusively for the right (left) eigenvector of $\T_{\tau_\lambda}$
and write $\ket{r_\lambda, s}$ ($\bra{l_\lambda,s}$) for the right (left) eigenvectors of $\T_\tau$ for $\tau > \tau_\lambda$. 
We call such eigenvalues with $\tau_\lambda = \tau$ relevant at $\tau$ and denote the leading (largest) relevant eigenvalue by $\lambda_1$.
Furthermore, we denote the leading nontrivial eigenvalue of $\T_\tau$ by $\lambda_0$ giving
$|\lambda_1|\leq|\lambda_0|\leq 1$, where $1$ is the trivial eigenvalue.
Assuming no accidental degeneracies the eigenvalue $\lambda$ is $(\tau - \tau_\lambda + 1)$-fold
degenerate.
The projection $\Proj_{\lambda, \tau}$ for given $\tau$ onto the corresponding
eigenspace can be constructed as follows:
For each $\rho \leq \tau$ we can choose the left and right eigenvectors 
corresponding to fixed $\tau_\lambda=\rho$ to be biorthogonal, i.e. $\braket{l_\lambda|r_{\lambda'}}=\delta_{\lambda,\lambda'}$.
This guarantees that the vectors $\bra{l_\lambda, s}$, $\ket{r_{\lambda'}, s}$ are biorthogonal, i.e.,
$\braket{l_\lambda, s | r_{\lambda'}, s'} = \delta_{s,s'}\delta_{\lambda,\lambda'}$.
The projections onto the corresponding eigenspaces are given by
\begin{align}
 \Proj_{\lambda, \tau} = \sum_{s = 0}^{\tau - \tau_\lambda} \ket{r_\lambda, s}\!
\bra{l_\lambda, s}   
\label{eq:proj_dual_unitary}
\end{align}
for nontrivial eigenvalues and $\Proj_{1, \tau} = \ket{\circ}\!\bra{\circ}^{\otimes \tau}$.
They form -- using the numerically observed fact that $\T_\tau$ is diagonalizable --
a resolution of identity
$\sum_\lambda \Proj_{\lambda, \tau} = \mathds{1}_{q^{2\tau}}$.

\begin{figure}[]
    \includegraphics[width=8.5cm]{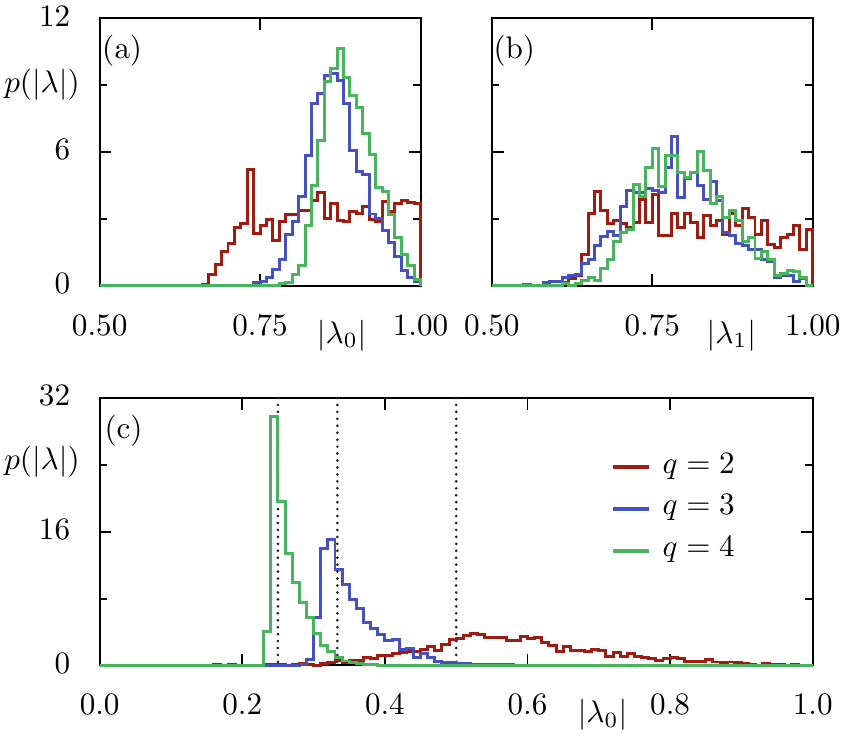}
    \caption{Distribution $p(|\lambda|)$ of (a) the largest nontrivial eigenvalue $\lambda_0$ and (b) the largest relevant eigenvalue $\lambda_1$  for T-dual impurity interactions for $q=2,3,4$  (with corresponding $\tau$ given by (a) $\tau=10,6,4$ and (b) $\tau=6,4,3$). Panel (c) depicts the distribution $p(|\lambda|)$ of the largest nontrivial eigenvalue $\lambda_0$ for generic impurity interactions for $q=2,3,4$ (with corresponding $\tau$ given by $\tau=10,6,4$). Dotted lines correspond to $|\lambda_0|=1/q$. All histograms are created from $>1000$ realizations with 
    Haar random $U$ in the generic case, and $U$ with Haar random local unitaries $u_\pm,v_\pm$ and fixed $J=1/2$ in the T-dual case; see Eq.~\eqref{supp_eq:parametrization_T_dual}.}
    \label{fig:leading_eigenvalues_distribution}
\end{figure}

Writing $\T_\tau = \sum_\lambda \lambda \Proj_{\lambda, \tau}$ and inserting into
Eq.~\eqref{eq:correlations_via_transfer_matrix} we obtain for $t=L\tau + \delta$
\begin{align}
    C_{ab}(t)
    = \sum_{\lambda, \sigma} \lambda^{L - \delta}\sigma^{\delta} 
    (\bra{l_\lambda}\!\otimes\!\bra{b})\ket{r_\sigma}\!
    \bra{l_\sigma}(\ket{a}\!\otimes\!\ket{r_\lambda})
    \label{eq:correlations_from_spectral_decomposition}\,,
\end{align}
where the sums run over all nontrivial eigenvalues $\lambda \in \text{spec}(\T_\tau)$
and $\sigma \in \text{spec}(\T_{\tau + 1})$ for which $\tau_\lambda = \tau$ and 
$\tau_\sigma = \tau + 1$.
The latter restriction is due to the property that $\tr\left(\left[\Proj_{\lambda, \tau} \otimes \mathds{1}_{q^2}\right]\Proj_{\sigma, \tau + 1}
\C_{ab, \tau + 1}\right) = 0$ if $\tau_\lambda < \tau$ or $\tau_\sigma < \tau + 1$, essentially following from $\tr(a)=\tr(b) =0$; see App.~\ref{S-V} for details.
This justifies the notion of relevant eigenvalues, as 
only the eigenvalues relevant at $\tau$ and $\tau + 1$ contribute to the correlation functions~\eqref{eq:correlations_via_transfer_matrix_diagramm}.
Their asymptotic scaling is hence determined by the leading relevant
eigenvalues of $\T_\tau$ and $\T_{\tau + 1}$, respectively.
Assuming unique leading relevant eigenvalues $\lambda_1$ and $\sigma_1$, the correlations scale as
\begin{align}
    C_{ab}(t) \sim \lambda_1^{L - \delta} \sigma_1^\delta 
    (\bra{l_{\lambda_1}}\!\otimes\!\bra{b})\ket{r_{\sigma_1}}\!
    \bra{l_{\sigma_1}}(\ket{a}\!\otimes\!\ket{r_{\lambda_1}})
    \label{eq:correlations_asymptotics}
\end{align}
if both $L - \delta$ and $\delta$ are large.
Here the factors involving the scalar products of eigenvectors as well as initial and final operators
depend on $\tau$ only but not on $L$ and $\delta$.
Hence for fixed $\tau\ge 1$ and arbitrary $\delta$ correlations are bounded by
$|C_{ab}(t)| \le \text{const}\times \left(\max\{|\lambda_1|, |\sigma_1|\}\right)^L$, implying exponential suppression of correlations in large but finite systems for $t>L$,
provided there is a spectral gap, i.e. $|\lambda_1|<1$, $|\sigma_1|<1$.
By numerically computing both the leading nontrivial eigenvalue $\lambda_0$ and the leading
relevant eigenvalue $\lambda_1$ for the respective largest accessible system sizes, we confirm that, with probability one, the leading (relevant) eigenvalues have modulus smaller than one.
For this we choose more than 1000 realizations with fixed $J=1/2$ and Haar random $u_\pm, v_\pm \in \text{U}(q)$, see Eq.~\eqref{supp_eq:parametrization_T_dual}.
The corresponding probability densities $p(|\lambda|)$ are depicted in
Fig.~\ref{fig:leading_eigenvalues_distribution}(a,b).
For $q \in \{3, 4\}$ the probability density $p(|\lambda|)$ approaches zero for $|\lambda| \to 1$ for both the leading nontrivial and leading relevant eigenvalue.
Although this is not the case for qubits, $q=2$, we found no instance for which $|\lambda_1| = 1$.
This difference between qubits, $q=2$, and $q\geq3$ might be related due the fact that T-dual impurity interactions cannot be maximally entangling as their entangling power is strictly smaller than the maximal possible value \cite{Zan2001,AraRatLak2021}.
In contrast, for larger $q$ the impurity interaction may exhibit the maximal possible entangling power.
Further note that $p(|\lambda|)$ depends only weakly on $\tau$, i.e. the 
distributions depicted in Fig.~\ref{fig:leading_eigenvalues_distribution} do not change significantly with $\tau$.
We therefore conclude that the exponential suppression of correlation functions with system size is a
generic feature of the boundary-chaos model in case of T-dual impurities.

\begin{figure}[]
    \includegraphics[width=8.5cm]{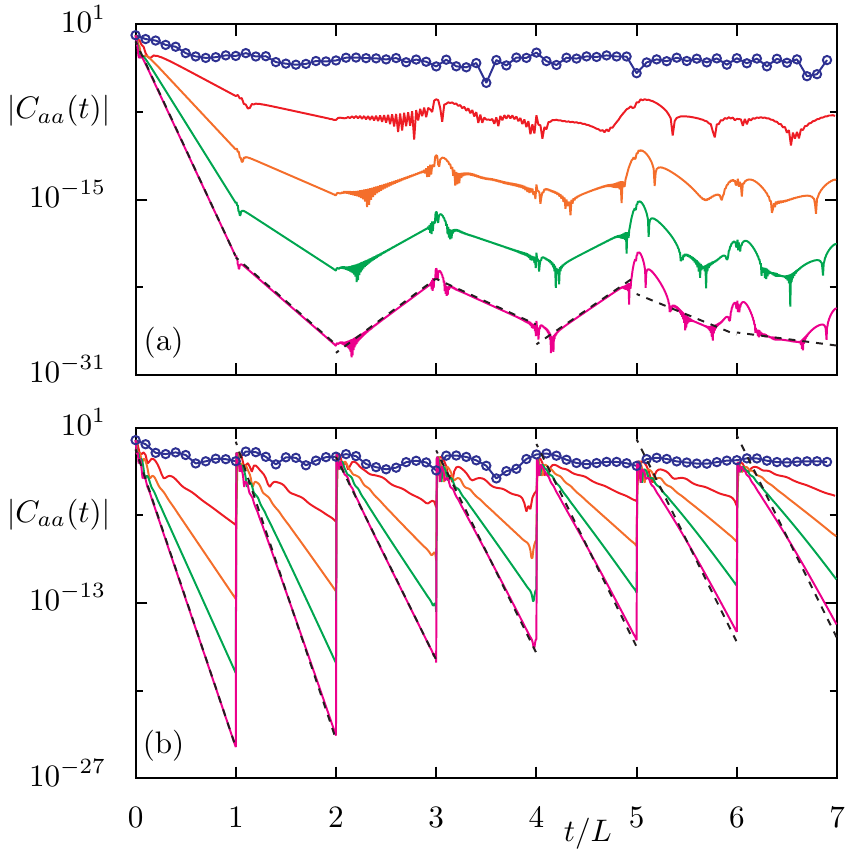}
    \caption{Autocorrelation functions $C_{aa}(t)$ for (a) T-dual and (b) generic impurities for qubits, $q=2$, and $a=\sigma_z$ via Eq.~\eqref{eq:correlations_via_transfer_matrix}. Lines correspond to  system sizes (top to bottom) $L=10$ (blue), $L=50$ (red), $L=100$ (orange), $L=150$ (green), and 
        $L=200$ (magenta). Blue circles denote correlation functions obtained from exact diagonalization at $L=10$. Dashed lines depict the asymptotic scaling from (a) Eq.~\eqref{eq:correlations_asymptotics} and (b) $\sim \sigma_0^\delta$ with fitted prefactor.}
    \label{fig:correlations}
\end{figure}

This is also depicted in Fig.~\ref{fig:correlations}(a), where correlations for a representative
example system are shown for system sizes between $L=10$ and $L=200$ clearly demonstrating the
exponential dependence of correlations on $L$.
In particular for $L=200$ we find good qualitative agreement with the asymptotic
result~\eqref{eq:correlations_asymptotics}.
Here the computationally maximal accessible system size is restricted by machine precision due to the exponential suppression.
In particular for sufficiently small leading relevant eigenvalues correlations for much larger system sizes than $L=200$ can be obtained.
Time intervals in which correlations decay with $t$ correspond to $\tau$ for which 
$|\lambda_1|>|\sigma_1|$ while growing correlations correspond to
$|\sigma_1|>|\lambda_1|$.
Corrections to the asymptotic
scaling are dominated by the next
to leading relevant eigenvalues $\lambda_2$ and $\sigma_2$ of $\T_\tau$ and $\T_{\tau + 1}$ and are
of the order $\left(\lambda_2/\lambda_1\right)^{L-\delta}
\left(\sigma_2/\sigma_1\right)^{\delta}$.
Note that the gap between the leading and subleading eigenvalue will in general shrink as $\tau$ grows.
In any case deviations to the asymptotic scaling are most prominent when $t \approx \tau L$, i.e., when either $\delta$ or $L-\delta$ are small as it is also seen in Fig.~\ref{fig:correlations}(a). \\

\begin{figure}[]
    \includegraphics{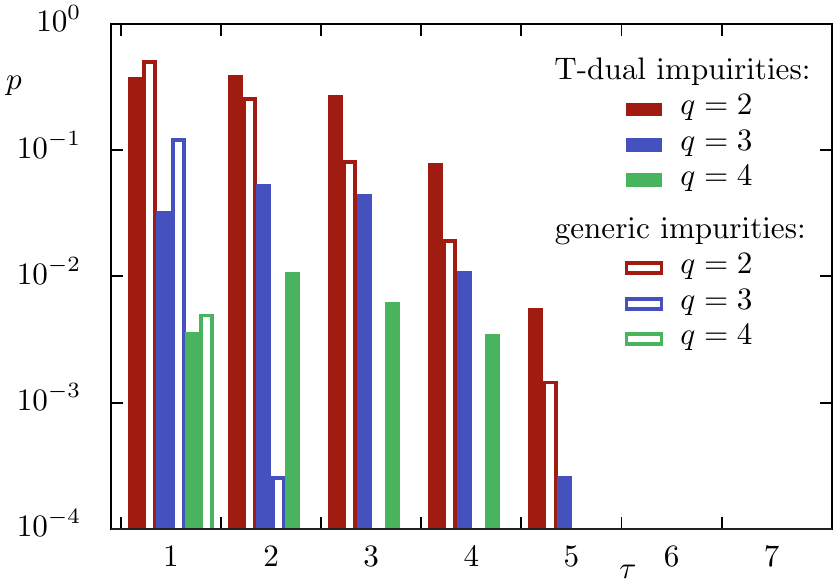}
    \caption{Probability $p(|\lambda_0(\tau + 1)| > |\lambda_0(\tau)|)$, i.e., for the largest nontrivial
        eigenvalue to grow when advancing from $\tau$ to $\tau + 1$ for various $q$ and for both
        the T-dual and the generic case. Here we use more than 2000 realizations and 500 for the largest accessible values of $\tau$, respectively from the same ensembles as used for Fig.~\ref{fig:leading_eigenvalues_distribution}. The maximum system size is given by $\tau+1=11$, $\tau+1=8$, and $\tau+1=5$ for $q=2, 3, 4$, respectively.
        In particular, for $q=2, 3$ the probability is found to be zero, when there is no bar depicted. For $q=4$ this holds only up to $\tau=4$ as we compute the leading eigenvalue only up to $\tau+1=5$.}
    \label{fig:leading_eigenvalues}
\end{figure}

In order to establish exponential suppression of correlations with system size for all $\tau$ one
needs to show that there is a spectral gap $\Delta_{1} = \limsup_{\tau \to \infty}
\left(1 - |\lambda_1(\tau)|\right) > 0$, where
$\lambda_1(\tau)$ denotes the leading relevant eigenvalue at $\tau$.
We are able to address this question only numerically by  considering instead the leading nontrivial eigenvalue $\lambda_0(\tau)$ of $\T_{\tau}$ and the corresponding spectral gap  $\Delta_{0} = \lim_{\tau \to \infty}
\left(1 - |\lambda_0(\tau)|\right)$, 
as $\lambda_0(\tau)$ can be computed efficiently using Arnoldi iteration and the matrix-product structure of $\T_\tau$.
Note that $|\lambda_0(\tau)|$ grows monotonically with $\tau$.
For T-dual impurities drawn from the ensemble defined above, i.e., fixed $J=1/2$ and Haar random local unitaries, we find the probability $p(|\lambda_0(\tau +1)|>|\lambda_0(\tau)|)$ for the leading eigenvalue to grow from $\tau$ to $\tau + 1$ to quickly decrease with $\tau$.
This is illustrated in Fig.~\ref{fig:leading_eigenvalues}, where we depict $p(|\lambda_0(\tau +1)|>|\lambda_0(\tau)|)$ as a function of $\tau$.
Thus the leading eigenvalue will generically be constant from a fixed $\tau$ on.
Combining this with the properties of the probability distribution for the leading eigenvalues we conclude that for generic choices of T-dual impurities there exists a finite spectral gap 
$\Delta_{0}>0$ and hence $\Delta_{1}>0$ implying exponential suppression of correlation
functions~\eqref{eq:correlation} with system size for any $\tau = \lfloor t/L\rfloor \ge 1$.
However, due to the overlaps between eigenstates and initial and final operators in 
Eq.~\eqref{eq:correlations_from_spectral_decomposition} this might not be uniform in $\tau$.

\section{Generic unitary impurities}
In the following we comment on the case of generic unitary impurity interactions $U \in \text{U}(q^2)$ for which correlation functions exhibit qualitatively different properties.
This again can be understood in terms of the spectral properties of the transfer matrices
$\T_{\tau}$, which differ in some important points from the T-dual case.
We list those differences obtained using both numerical and analytical arguments below.
For more details we refer the reader to App.~\ref{S-VI}.
Firstly, $\T_\tau$ fails to be diagonalizable but exhibits nontrivial Jordan blocks, which for eigenvalue $\lambda$ is of dimension $\tau - \tau_\lambda + 1$.
Secondly, using the notation of Eq.~\eqref{eq:eigenvectors}, only $\bra{l_\lambda, 0}$ and $\ket{r_\lambda, \rho - \tau}$ is a left or, respectively, right eigenvector of $\T_\rho$, implying $\text{spec}(\T_\tau)\subseteq \text{spec}(\T_{\tau+1})$ also for generic impurity interactions.
We have to consider the Jordan decomposition of the transfer matrices, for which the projection onto the Jordan block corresponding to $\lambda$ is not given by Eq.~\eqref{eq:proj_dual_unitary} but can only be constructed numerically for small $\tau$.
Inserting the Jordan decomposition into Eq.~\eqref{eq:correlations_via_transfer_matrix} nevertheless yields the asymptotic scaling of correlations
with system size.
In contrast to the T-dual case now the notion of relevant eigenvalues breaks down as all
pairs $\lambda,\sigma$ of eigenvalue $\lambda$  of $\T_\tau$ and nontrivial eigenvalue $\sigma$ of $\T_{\tau + 1}$, 
contribute to the correlation function.
Note that the trivial eigenvalue $1 \in \text{spec}(\T_\tau)$ is not excluded and in the absence of
further eigenvalues with modulus one dominates the correlation function for $L-\delta$ being large.
Replacing $\T_\tau$ by the projection onto the trivial eigenspace $\ket{\circ}\!\bra{\circ}^{\otimes \tau}$ and reversing the mapping from the original tensor network~\eqref{eq:correlations_folded} to the helical network~\eqref{eq:correlations_via_transfer_matrix_diagramm} allows for some intuition on the role of the trivial eigenvector:
The local operator $a$ cannot travel along the boundary for more than $\delta$ subsequent time steps before being scattered into the bulk, where it undergoes free dynamics.
This minimizes the total number of scattering events at the impurity interaction from $t$ to $\tau \times \delta$ and hence gives rise to the dominant contribution to the correlation function.
In contrast, the above process has a vanishing contribution to the correlation function in the T-dual case as a consequence of dual unitarity of the folded gates $V$.

In order to estimate contributions from the remaining part of the spectrum we confirm that generically the leading nontrivial eigenvalue has modulus smaller 
than one by computing the leading 
nontrivial eigenvalue $\lambda_0$ for the largest accessible $\tau$ numerically for Haar-random impurities.
The resulting probability distribution $p(|\lambda_0|)$ is shown in
Fig.~\ref{fig:leading_eigenvalues_distribution}(c). The probability density is peaked at $1/q$ and approaches $0$ as
$|\lambda_0| \to 1$. 
Moreover, we expect nonzero spectral gap $\Delta_0$, i.e., the leading nontrivial eigenvalue obeys $|\lambda_0(\tau)|<1$  for all $\tau$.
This is supported by the decrease of the probability $p(|\lambda_0(\tau + 1)| > |\lambda_0(\tau)|)$ of the leading nontrivial eigenvalue to grow from $\tau$ to $\tau+1$ with $\tau$ which we illustrate in Fig.~\ref{fig:leading_eigenvalues}.

Hence the asymptotic large $L$ scaling of correlation functions is obtained by replacing $\T_\tau$ in Eq.~\eqref{eq:correlations_via_transfer_matrix} by the projection
$\ket{\circ}\!\bra{\circ}^{\otimes \tau}$ onto the eigenspace corresponding to the trivial eigenvalue.
This results in
\begin{align}
C_{ab}(L\tau+\delta) & \sim \bra{\circ}^{\otimes \tau}\!\otimes\!\bra{b}\T_{\tau + 1}^{\delta}\ket{a}\!\otimes\!\ket{\circ}^{\otimes \tau} 
\label{eq:asymptotic_correaltion_generic_1}\\
& =\, \begin{tikzpicture}[baseline=32, scale=0.6]
\foreach \x in {0,...,3}
\foreach \y in {1,2,3}
{
    \transfermatrixgate{\x}{\y}
}
\foreach \y in {1,2,3}
{
    \draw[thick] (-0.5, \y) -- (- 0.7, \y);
    \draw[thick] (3.5, \y) -- (3.7, \y);    
    \IdState{-0.7}{\y}
    \IdState{3.7}{\y}
}
\foreach \x in {0, 1,2, 3}
{
    \draw[thick] (\x, 3.5) -- (\x, 3.7);
    \draw[thick] (\x, 0.5) -- (\x, 0.3);    
}
\foreach \x in {0, 1, 2}
{
    \IdState{\x}{3.7}
    \IdState{\x + 1}{0.3}
}
\localOperator{3}{3.7}
\localOperator{0}{0.3}
\Text[x=-0.3,y=0.1]{$a$}
\Text[x=3.3,y=3.8]{$b$}
\draw [thick, stealth-stealth](4.25, 3.15) -- (4.25, 0.85);
\Text[x=4.5,y=2.]{\footnotesize$\delta$}
\draw [thick, stealth-stealth](-0.2, -0.25) -- (3.2, -0.25);
\Text[x=1.6,y=-0.6]{\footnotesize$\tau + 1$}
\end{tikzpicture}. \label{eq:asymptotic_correaltion_generic_2}
\end{align}
which coincides with spatiotemporal correlation functions 
$\braket{b_{x+r}|a_x(t)}$ in constant gate brickwork quantum circuits, see e.g.~\cite{KosBerPro2021}, for $t-r=\delta$, $t+r=\tau+1$, 
and is in general nonzero. 
For large $\delta$, Eq.~\eqref{eq:asymptotic_correaltion_generic_1} is dominated by the leading nontrivial eigenvalue $\sigma_0$ of $\T_{\tau + 1}$ as the trivial eigenvector is orthogonal to both 
$\ket{a}\otimes\ket{\circ}^{\otimes \tau}$ and $\ket{\circ}^{\otimes \tau}\otimes\ket{b}$.
Consequently, $C_{ab}(L\tau+\delta) \sim \sigma_0^\delta$ independent from $L$ as long as 
$L-\delta \gg \tau$ and $\delta \gg \tau$, i.e., larger than the largest Jordan blocks.
This implies persistent revivals of correlation functions with period $L$ to a value which is approximately independent of the system size.
Thus for generic impurity interactions correlations are not exponentially
suppressed in $L$ for all times.
For a representative example this is depicted in Fig.~\ref{fig:correlations}(b) showing clear 
signatures of
persistent revivals.
Nevertheless, the leading nontrivial eigenvalue of $\T_{\tau + 1}$ yields the correct asymptotic 
scaling as it
is indicated by the dashed black lines for $L=200$ with a fitted prefactor.
Deviations from the asymptotic scaling $\sim \sigma_0^\delta$ are most prominent around $t \approx 
\tau L$, i.e., 
for small or large $\delta$, and are due to both the 
subleading parts of the spectra of $\T_\tau$ and $\T_{\tau + 1}$
as well as the nontrivial Jordan structure.
Nevertheless, by evaluating the diagrammatic expression~\eqref{eq:asymptotic_correaltion_generic_2} for large $\tau$ and small (fixed) $\delta$
in terms of column transfer matrix (with leading nontrivial eigenvalue $\chi$, $|\chi|<1$) instead of the row transfer matrices $\T_{\tau + 1}$ we find the dominant contribution $\sim \chi^\tau$. This suggest exponential correlation decay on an extensive time scale $-L/\log|\chi|$, which is expected due to the $1/L$ density of impurities. 
In particular, this causes the amplitude of the revivals to decay exponentially with $\tau$ and the system will eventually thermalize.

\section{Correlation functions for arbitrary local operators \label{sec:generic_locations}}

The techniques to study correlations functions between local operators $a_0$ and $b_0$ acting on the first lattice site can be extended to local operators acting on arbitrary sites.
To this end let $a_x$ and $b_y$ denote operators acting as traceless Hermitian operators $a$ and $b$ on lattice sites $x$ and $y$, respectively, and trivially otherwise.
Again we are interested in correlation functions given by  $N^{-1}\tr\left(\U^{-t}a_x\U^tb_y\right)$.
Due to unitality of the folded gates $W$ and the free dynamics in the bulk it is sufficient to consider only local operators $a_x$ acting on the lattice sites $x\in\{0,1\}$ and $b_y$ acting on sites $y\in\{0,2\}$.
We denote their correlations by 
\begin{align}
    D_{ab}(t) = \frac{1}{N}\tr\left(\U^{-t} a_x \U^t b_y\right).
    \label{eq:correlations_diff_sites}
\end{align}
Dynamical correlations for different lattice sites are just shifted in time by  $\Delta t = \Delta t_x + \Delta t_y \leq 2(L-1)$ and are zero for times $t<\Delta t$.
Here, $\Delta t_x = L -x/2$ if $x$ is even, $\Delta t_x = (x-1)/2$ if $x$ is odd, and $\Delta t_x=0$ if $x=0$, whereas $\Delta t_y = y/2 - 1$ if $y$ is even, $\Delta t_y = L - (y+1)/2$ if $y$ is odd, and $\Delta t_y = 0$ if $y=0$.
Hence we focus on the computation of $D_{ab}(t)$ and 
for the sake of concreteness consider only the case $x\neq 0 \neq y$ in the following.
Writing again $t=\tau L + \delta$ we obtain a tensor network representation of $D_{ab}(t)$ as
\begin{align}
D_{ab}(t) &  =
\begin{tikzpicture}[baseline=32, scale=0.6]
 \foreach \x in {0,...,3}
{
    \draw[thick] (\x, -0.5 ) arc (180:360: 0.15);
    \draw[thick] (\x, 4.5 ) arc (360:180: -0.15);
    \draw[thick, color=gray, dashed] (\x + 0.3, -0.5) -- (\x + 0.3, 4.5);    
}
\foreach \x in {0,...,2}
{
    \Shift{\x+0.5}{0.}
}
\foreach \x in {0,...,3}
\foreach \y in {1,2,3}
{
    \transfermatrixgate{\x}{\y}
}
\foreach \y in {1,2, 3}
{
    \draw[thick] (-0.5, \y) -- (- 0.7, \y);
    \draw[thick] (3.5, \y) -- (3.7, \y);    
    \IdState{-0.7}{\y}
    \IdState{3.7}{\y}
}
\foreach \x in {0,...,2}
\foreach \y in {4}
{
    \transfermatrixgate{\x}{\y}
}
\Identity{3.}{4.}
\foreach \y in {4}
{
    \draw[thick] (-0.5, \y) -- (- 0.7, \y);
    \draw[thick] (2.5, \y) -- (2.7, \y);    
    \IdState{-0.7}{\y}
    \IdState{2.7}{\y}
}
\localOperator{3.7}{3}
\localOperator{-0.7}{1}
\Text[x=-0.8,y=1.-0.4]{$a$}
\Text[x=3.8,y=3.+0.4]{$b$}
\initialOperatorId{-0.5}{0.}
\finalOperatorId{3.5}{0.}
\draw [thick, stealth-stealth](4.2, 3.15) -- (4.2, 0.85);
\Text[x=4.45,y=2.]{\footnotesize$\delta$}
\draw [thick, stealth-stealth](4.9, 4.15) -- (4.9, 0.85);
\Text[x=5.15,y=2.5]{\footnotesize$L$}
\draw [thick, stealth-stealth](-0.2, -0.8) -- (3.2, -0.8);
\Text[x=1.6,y=-1.05]{\footnotesize$\tau + 1$}
\end{tikzpicture}. \label{eq:correlations_via_transfer_matrix_diff_sites} 
\end{align}

\begin{figure}[t]
    \includegraphics[width=8.5cm]{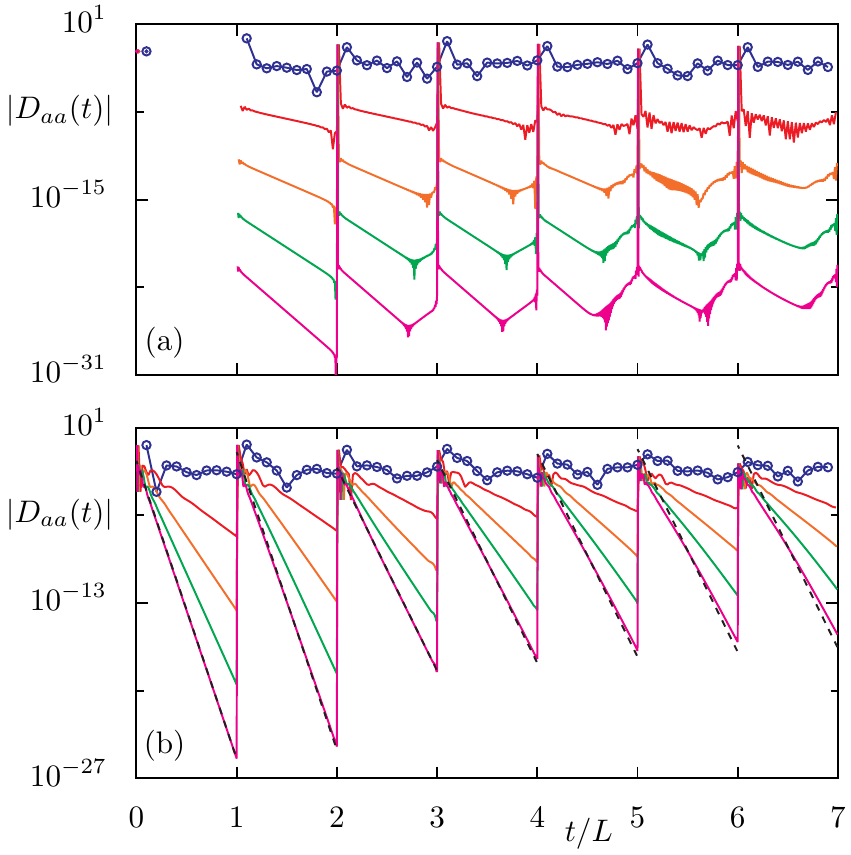}
    \caption{ Correlation functions $D_{aa}(t)$ for (a) T-dual and (b) generic impurities for qubits, $q=2$, and $a=\sigma_z$ via Eq.~\eqref{eq:correlations_via_transfer_matrix} in the same systems as in Fig.~\ref{fig:correlations}. Lines correspond to  system sizes (top to bottom) $L=10$ (blue), $L=50$ (red), $L=100$ (orange), $L=150$ (green), and 
        $L=200$ (magenta). Blue circles denote correlation functions obtained from exact diagonalization at $L=10$. (b) Dashed lines depict the asymptotic scaling $\sim \sigma_0^\delta$ with fitted prefactor. For T-dual impurity interactions $D_{aa}(t)=0$ for $1 < t < L+1$.}
    \label{fig:correlations_different_sites}
\end{figure}

Structurally, the above network is the same as the network~\eqref{eq:correlations_via_transfer_matrix_diagramm}, with only $\C_{ab}$ replaced by $\C_{\mathds{1} \mathds{1}}$, the left boundary conditions of the bottom most transfer matrix $\T_{\tau + 1}$ replaced by $a$, and the right boundary condition of the top most transfer matrix $\T_{\tau + 1}$ replaced by $b$.
Consequently, for large $L$ and large $\delta$ the correlations are determined by the spectral properties of transfer matrices $\T_\tau$ and $\T_{\tau+1}$ with the original boundary conditions given by $\ket{\circ}$.
Replacing these transfer matrices by their spectral or Jordan decomposition respectively allows for studying the asymptotics of correlations. \\

Again, we first consider the case of T-dual impurity interactions.
T-duality causes correlations to vanish for times $1<t<L$, i.e.
$\tau=0$ and $\delta>0$.
For $\tau>0$ the notion of relevant eigenvalues as defined for $x=y=0$ breaks down.
Instead, for  $\delta \neq 1$, the pairs of nontrivial eigenvalues $\lambda$ of $\T_\tau$ and $\sigma$ of $\T_{\tau + 1}$ with $\tau_\sigma = \tau_\lambda + 1$ contribute to the correlations.
Each such pair gives a contribution  $\sim \lambda^{L-\delta}\sigma^{\delta}$ to the correlation function.
Assuming a spectral gap $\Delta_0 > 0$ this gives rise to exponential suppression of correlations.
This is depicted in Fig.~\ref{fig:correlations_different_sites}(a), 
where correlations clearly show exponential suppression for $\delta\neq 1$.
The asysmptotic scaling, however, at fixed $\tau$ is typically not given by a single pair of eigenvalues, as pairs for which 
$\sim \lambda^{L-\delta}\sigma^{\delta}$ decreases with $\delta$ compete with pairs for which this expression grows with $\delta$, leading to the switching between positive and negative slope of the correlations at fixed $\tau$.
In contrast to the above, for $\delta=1$ the correlations are
approximately independent from system size leading to persistent $L$-periodic revivals.
These revivals are due to the trivial eigenvalue of $\T_\tau$, which gives a non-vanishing contribution to the correlation function for $\delta=1$.
Instead the correlations are determined by a two-qudit quantum channel $\mathcal{M}$ introduced in Ref.~\cite{BerKosPro2019}, where it governs correlations along light rays in dual-unitary brickwork quantum circuits.
The asymptotic correlation functions then read $\bra{b}\mathcal{M}^\tau\ket{a}$.
Hence the amplitude of the revivals decays with $\tau$ as $\chi^\tau$ with $\chi$ the leading nontrivial eigenvalue of $\mathcal{M}$, implying thermalization on an extensive time scale $-L/\log|\chi|$ for typical T-dual impurities.
In contrast if either of the local operators $a_x$ or $b_y$ acts on lattice site $x=0$ or $y=0$, respectively, revivals of correlations do not appear and correlations are exponentially suppressed at all times. \\

For generic impurity interactions the analysis of correlation functions is up to minor details the same for $D_{ab}(t)$, Eq.~\eqref{eq:correlations_diff_sites}, and $C_{ab}(t)$, Eq.~\eqref{eq:correlation}.
That is, for large $L$ the correlations are dominated by the trivial eigenvector of $\T_\tau$ leading to persistent revivals of correlations with period $L$, which decay on an extensive time scale as discussed in the case of $x=y=0$.
For both $L$ and $\delta$ large, correlations scale as $\sigma_0^{\delta-2}$ with $\sigma_0$ the leading nontrivial eigenvalue of $\T_{\tau+1}$.
The correlations and their asymptotic scaling are depicted in Fig.~\ref{fig:correlations_different_sites}(b) and show qualitatively the same behavior as in Fig.~\ref{fig:correlations}(b) for $x=y=0$. \\

\section{Conclusions}
The concept of boundary chaos provides a minimal model of an interacting many-body systems, which despite exhibiting quantum chaos allows for a rigorous treatment via the mapping to the helical circuit.
This effectively reduces the dimension of the relevant tensor networks $t\times L \to [t/L]\times L = t$.
We use this mapping to efficiently compute correlation functions between local operators for system sizes much larger than what is accessible by direct computation for times up to a few multiples of system size.
For impurity interactions fulfilling the T-duality property we establish exponential suppression of correlations with system 
size for all times and for local operators acting nontrivially at the first lattice site.
For arbitrary locations of the operators this is still the case for almost all times as $L \to \infty$.
In contrast, for generic impurity interactions correlations show persistent revivals with a period given by the system size $L$ irrespective of the location of the operators.
The amplitude of these revivals nevertheless decays exponentially with $\tau$ leading to thermalization on an extensive (in $L$) time scale.
Our construction bears curious analogy with the method of Poincar\'e section in few-body dynamics~\cite{Ott2002}, specifically with quantized Poincar\' e transfer operator~\cite{Bog1992,Pro1995} as well as with the influence matrix approach to many-body quantum dynamics~\cite{LerSonAba2021}. Our method `integrates out' the free flights through swap circuits between subsequent impurity interactions and hence is able to exactly describe the influence of the trivial bulk on the nontrivial boundary.
As swap gates represent a particular solution of Yang-Baxter equation, one may expect that our concept can be extended to boundary perturbations of generic integrable circuits where one would need to `integrate out' interacting integrable dynamics between collisions.
 Aside from a natural extension of our analysis to operator entanglement and out-of-time-order correlations, it may allow for a rigorous proof of ETH once the existence of a spectral gap can be established.
Moreover, one should attempt a rigorous analysis of the spectral form factor, which should reveal more structure than in uniform dual-unitary circuits~\cite{BerKosPro2021}, e.g. we observe a nontrivial Thouless time scale; see App.~\ref{S-II}. \\

\section*{Acknowledgements}

The work has been supported by Deutsche Forschungsgemeinschaft (DFG), Project No. 453812159 (FF), and by European Research Council (ERC), Advanced grant 694544-OMNES, and by Slovenian Research Agency (ARRS), under program P1-0402 (TP). We thank B. Bertini and P. Kos for discussions and valuable remarks on the manuscript and M. Mestyan for collaboration in the preliminary stage of this project. 

\appendix

\begin{widetext}

\HIDDEN{In the following we present some additional data, and provide detailed derivation of all the results reported in the main text. It is divided into six sections.

In section \ref{S-I} we review the parametrization of T-dual gates which is used in this paper and which provides a complete parametrization of T-dual gates of qubits. In \ref{S-II} we report on numerical investigation of spectral statistics in the models of boundary chaos, as studied from the perspective of correlation functions in the main text, corroborating on their quantum chaotic nature. In section \ref{S-III} we provide fully detailed boundary-to-helix mapping of our circuits, in particular, we construct an isomorphism between propagators on the two lattice domains. In section \ref{S-IV} we provide some numerical analysis of the spectra of transfer matrices for both cases of T-dual and generic impurity gates. Particularly valuable is the analysis of the growth (or no-growth) of the leading nontrivial eigenvalue with scaled time $\tau = \lfloor t/L\rfloor$. In section \ref{S-V} we formally characterise the structure of the eigenstates in the T-dual case, which allow us to make precise conclusions on the decay of correlations. Finally, in section \ref{S-VI} we discuss the Jordan structure of the transfer matrices in the generic case and their consequence on the asymptotic decay of correlations.}

\section{Parameterization of T-dual Impurities}
\label{S-I}

A unitary matrix $U \in \text{U}(q^2)$ acting on $\mathds{C}^q \otimes  \mathds{C}^q$ 
is called T-dual~\cite{AraRatLak2021}, if its partial transpose $U^{\text{T}_1}$ (with respect to the first tensor factors) is unitary as well.
More precisely, denoting the matrix elements with respect to the canonical product basis by
$U^{ij}_{kl}$, the matrix elements of the partial transpose are given by
\begin{align}
    \left(U^{\text{T}_1}\right)^{ij}_{kl} = U^{kj}_{il}.
\end{align}
Equivalently, one might define T-duality by requesting that the partial transpose with respect to the second tensor factor is unitary, as unitarity under partial transpose with respect to one tensor factor implies unitarity under partial transpose with respect to the other tensor factor.
A parametrization of a subset of T-dual gates can be obtained from a parametrization of dual-unitary gates \cite{ClaLam2021,BerKosPro2021} (which is complete for $q=2$ \cite{BerKosPro2019}):
\begin{align}
    U_{\text{du}} = \left(u_+ \otimes u_-\right) \exp\left(\ui J \sigma_{q^2-1} \otimes \sigma_{q^2-1} \right) P \left(v_- \otimes v_+\right) \in \text{U}(q^2)
\end{align}
with $\sigma_i$ the generalized Gell-Mann matrices, $J \in \left[0, \pi/4\right]$ and
$u_\pm, v_\pm \in \text{U}(q)$.
Hence T-dual gates which are of the form $U = U_\text{du} P$ can be parameterized as given by Eq.~\eqref{supp_eq:parametrization_T_dual}
This parametrization, however, is neither surjective nor injective
for $q>2$.
For numerical investigations we fix $J=1/2$ throughout this article, and either pick a fixed set of generic local unitaries $u_\pm,v_\pm$ or sample them randomly and Haar distributed in ${\rm U}(q)$.
Note, that the operator entanglement entropy $E(U) = \sin^2(2J)/2$ \cite{Zan2001} 
as well as the entangling power $e(U)=2\sin^2(2J)/3$ \cite{AraRatLak2021} for $q=2$ is independent of the local unitaries.
In particular the latter is strictly smaller than the maximal value of $e(U)=2/3$ if $J<\pi/4$ and hence the ensembles of T-dual gates considered do not give rise to maximally entangling (chaotic) dynamics.

\section{Spectral Statistics}
\label{S-II}

\begin{figure}[]
    \includegraphics[width=\linewidth]{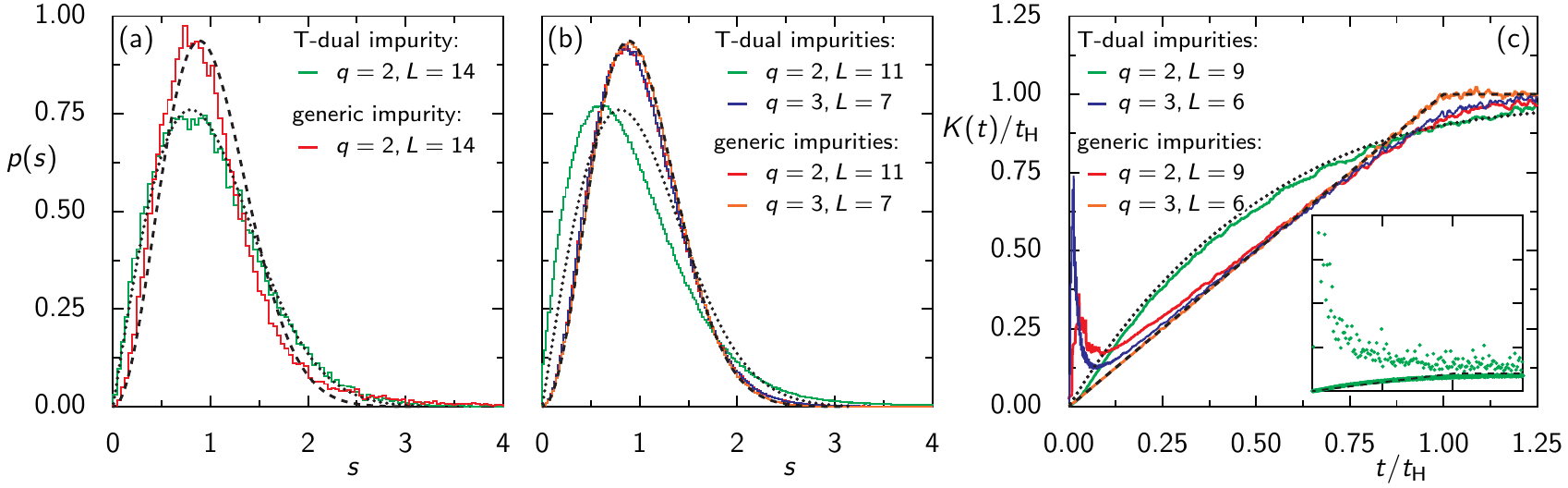}
    \caption{Level spacing distribution $p(s)$ for (a) the systems shown in Fig.~\ref{fig:correlations} and (b) for the ensemble average over Haar-Random generic impurities and T-dual impurities at fixed $J=1/2$ and Haar-random local unitaries $u_\pm$ and $v_\pm$, see Eq.~\eqref{supp_eq:parametrization_T_dual}. The dashed black line corresponds to the RMT result for the CUE and the dotted black line to the COE.
      Panel (c) depicts the normalized spectral form factor averaged over 1000 realizations of impurities. The data are further smoothed by a moving average over a window of 20 time steps. For the T-dual case with $q=2$ times $t=\tau L$ are excluded from the moving time average. The inset shows the spectral form factor for T-dual impurities with $q=2$ without smoothing. Black lines indicate the respective RMT spectral form factor with the dashed lines corresponding to the CUE and dotted lines corresponding to the COE. Time and $K(t)$ are scaled by the Heisenberg time $t_{\text{H}}=q^{L+1}$.}  
    \label{fig:level_spacing}
\end{figure}

In this section we present some numerical results on spectral properties of the
quantum circuits studied in the main text in order to justify our hypothesis, that the 
impurity interaction on the circuit's boundary is sufficient, despite non-interacting dynamics in the bulk, to generate random-matrix spectral 
statistics (and hence justify the name `boundary chaos'). Note that in a weaker context, namely considering local perturbation to integrable interacting spin chain, observation of quantum chaos has been reported in the literature before, see e.g.~\cite{San2004,BreLeBGooRig2020,Znidaric}.

To this end we consider two measures of level statistics: the nearest neighbor level-spacing distribution $p(s)$ and the spectral form factor $K(t)$. 
We find in general good agreement with random matrix theory (RMT) with the 
notable exception of T-dual impurities for qubits $q=2$.
In Fig.~\ref{fig:level_spacing}(a) we show the level-spacing distribution for the cases for which
we depict dynamical correlation functions in Fig.~\ref{fig:correlations} in the main text, where $q=2$.
At the chosen system size of $L=14$ the level-spacing distribution of the T-dual impurity interaction resembles Wigner-Dyson distribution corresponding to the circular orthogonal ensemble (COE), indicating existence of an antiunitary symmetry of the circuit.
For the generic impurity interaction the distribution is close to that of the circular unitary
ensemble (CUE) but is slightly shifted to the left.
This seems to be reminiscent of a generic feature of the systems studied here: For small system size the distribution is shifted to the left but approaches the corresponding random matrix result for large enough systems, possibly beyond what can be studied by numerical exact diagonalization.
However, except for T-dual qubit ($q=2$) circuits, we find random matrix spectral statistics even for moderate system sizes.
This is indicated in Fig.~\ref{fig:level_spacing}(b) for ensembles of impurities and local Hilbert space dimensions $q=2$ and $q=3$.
We computed spectral statistics also for $q=4$ but do not show the corresponding data as they can not be distinguished from $q=3$ on the shown scale.
In order to improve statistics we computed level statistics for an ensemble of systems with randomly sampled impurity gates, while we checked that the histograms of level spacing distribution do not change significantly for fixed typical impurities (not shown).
In the T-dual case the parameter $J=1/2$ entering the parameterization~\eqref{supp_eq:parametrization_T_dual} is kept fixed while the local unitaries $u_\pm$ and $v_\pm$ are drawn independently from the Haar ensemble.
In the generic case impurities are choosen Haar random.
Except for qubits in the T-dual case the level spacing distribution is close to that of the CUE.
For T-dual qubits the distribution is closer to the COE but is shifted to the left.
The RMT result might be approached for larger system sizes. \\

As a further spectral indicator of quantum chaos we compute the spectral form factor $K(t)$, i.e., the Fourier transform of the connected two-point function of the spectral density. 
The spectral form factor is computed via $K(t)=\langle |\tr\left(\U^t\right)|^2\rangle$, where the brackets denote the average over the ensembles of systems with varying impurity gates as described above.
This averaging is now necessary as the spectral form factor is not self-averaging \cite{Pra1997}. 
In Fig.~\ref{fig:level_spacing}(c) we again show data for both T-dual and generic impurities for local Hilbert space dimension $q=2$ and $q=3$. There, both the spectral form factor and time $t$ are measured in units of Heisenberg time $t_{\text{H}}=q^{L+1}$.
We again computed the spectral form factor also for $q=4$ but omit showing the data as it yields nearly identical results as $q=3$.
For the sake of a clearer presentation we additionally perform a moving time average over a window of width $\Delta t =20$.
As it is the case for the level spacing distribution except for T-dual qubits we find good agreement with the spectral form factor of the CUE given by $K(t)=t$ for $t<t_{\text{H}}$ and $K(t)=t_{\text{H}}$ for $t>t_{\text{H}}$ after some initial non-universal regime corresponding to the so-called Thouless time $t_{\text{Th}}$.
Remarkably, for generic impurity interactions and $q \geq 3$ we find almost no deviation from RMT spectral form factor for all times, indicating
$t_{\text{Th}}=0$ for large $q$. This curious observation certainly justifies a separate study.
For T-dual qubits we find strong enhancement of $K(t)$ at times $t=\tau L$ as it is shown in the inset where no moving time average is performed.
For the moving time average in the main panel we neglect these times and rescale the Heisenberg time by a factor $(L-1)/L$ accordingly.
This leads to a spectral form factor close to, but slightly below, the spectral form factor of the COE given by $K(t)=2t - 2t^2/t_{\text{H}} + 2t^3/t_{\text{H}}^2 -\ldots$ \, .

\section{Boundary to Helix Mapping}
\label{S-III}

In this section we provide the proof for the equality claimed in 
Eq.~\eqref{eq:correlations_via_transfer_matrix}.
In fact, we prove a stronger statement by providing a matrix product representation of
integer powers of the evolution operator
\begin{align}
    \W & = 
     \begin{tikzpicture}[baseline=3, scale=0.425]
    \foreach \j in {0}
    {
        \boundarygate{0./\sqrtwo}{\j/\sqrtwo}
    }
    \foreach \j in {0}
    \foreach \i in {1, 2, 4}
    {
        \Swap{2*\i/\sqrtwo}{\j/\sqrtwo}
    }
    \foreach \j in {1}
    \foreach \i in {1, 2, 4}
    {
        \Swap{2*\i/\sqrtwo - 1/\sqrtwo}{\j/\sqrtwo}
    }
    \foreach \j in {1}
    \foreach \i in {0}
    {
        \IdentityII{\i/\sqrtwo - 0.5/\sqrtwo}{\j/\sqrtwo}
    }
    \foreach \j in {1}
    \foreach \i in {9}
    {
        \IdentityII{\i/\sqrtwo - 0.5/\sqrtwo}{\j/\sqrtwo}
    }
    \draw[thick, black] (-.5/\sqrtwo + 5.,0.5/\sqrtwo) -- (-.5/\sqrtwo + 5. + 0.25, 0.5/\sqrtwo + 0.25);    
    \draw[thick, black] (-.5/\sqrtwo + 5.,1.5/\sqrtwo) -- (-.5/\sqrtwo + 5. + 0.25, 1.5/\sqrtwo - 0.25);   
    \draw[thick, black] (-.5/\sqrtwo + 7.,0.5/\sqrtwo) -- (-.5/\sqrtwo + 7. - 0.25, 0.5/\sqrtwo - 0.25);
    \draw[thick, black] (-.5/\sqrtwo + 7.,-0.5/\sqrtwo) -- (-.5/\sqrtwo + 7. - 0.25, -0.5/\sqrtwo + 0.25);
    \node at (5.5/\sqrtwo, 1./\sqrtwo)[circle,fill,inner sep=0.4pt]{};
    \node at (5.5/\sqrtwo - 0.25, 1./\sqrtwo)[circle,fill,inner sep=0.4pt]{};
    \node at (5.5/\sqrtwo + 0.25, 1./\sqrtwo)[circle,fill,inner sep=0.4pt]{};
    \node at (5.5/\sqrtwo, 0./\sqrtwo)[circle,fill,inner sep=0.4pt]{};
    \node at (5.5/\sqrtwo - 0.25, 0./\sqrtwo)[circle,fill,inner sep=0.4pt]{};
    \node at (5.5/\sqrtwo + 0.25, 0./\sqrtwo)[circle,fill,inner sep=0.4pt]{};
    \node at (5.5/\sqrtwo, -1.1/\sqrtwo)[circle,fill,inner sep=0.4pt]{};
    \node at (5.5/\sqrtwo - 0.25, -1.1/\sqrtwo)[circle,fill,inner sep=0.4pt]{};
    \node at (5.5/\sqrtwo + 0.25, -1.1/\sqrtwo)[circle,fill,inner sep=0.4pt]{};
    \draw[thin, gray] (-.5/\sqrtwo,-1.1/\sqrtwo) -- (5./\sqrtwo,-1.1/\sqrtwo);
        \draw[thin, gray] ( 6./\sqrtwo,-1.1/\sqrtwo) -- (-.5/\sqrtwo + 9./\sqrtwo,-1.1/\sqrtwo);
    \foreach \i in {0,...,5}
    {
        \draw[thin, gray] (-.5/\sqrtwo + \i/\sqrtwo,-0.8) -- (-.5/\sqrtwo + \i/\sqrtwo,-1.1/\sqrtwo);
    }
    \foreach \i in {0,2, 4}
    {
    \Text[x=-0.5/\sqrtwo+\i/\sqrtwo,y=-1.5/\sqrtwo]{\tiny $\i$};
    }
    \foreach \i in {7,8, 9}
    {
    \draw[thin, gray] (-.5/\sqrtwo + \i/\sqrtwo,-0.8) -- (-.5/\sqrtwo + \i/\sqrtwo,-1.1/\sqrtwo);
    }
    \Text[x=-0.5/\sqrtwo+7/\sqrtwo,y=-1.5/\sqrtwo]{\tiny $L-2$};
    \Text[x=-0.5/\sqrtwo+9/\sqrtwo,y=-1.5/\sqrtwo]{\tiny $L$};
    \Text[x=5./\sqrtwo,y=-1.9/\sqrtwo]{\tiny $x$}
    \end{tikzpicture}
    = \begin{tikzpicture}[baseline=3, scale=0.425]
    \foreach \j in {1}
    {
        \transfermatrixgateunrotated{0./\sqrtwo}{\j/\sqrtwo}
    }
    \foreach \j in {0}
    \foreach \i in {0, 1, 2, 4}
    {
        \Swap{2*\i/\sqrtwo}{\j/\sqrtwo}
    }
    \foreach \j in {2}
    \foreach \i in {1, 2, 4}
    {
        \Swap{2*\i/\sqrtwo - 1/\sqrtwo}{\j/\sqrtwo}
    }
    \foreach \j in {2}
    \foreach \i in {0}
    {
        \IdentityII{\i/\sqrtwo - 0.5/\sqrtwo}{\j/\sqrtwo}
    }
    \foreach \j in {2}
    \foreach \i in {9}
    {
        \IdentityII{\i/\sqrtwo - 0.5/\sqrtwo}{\j/\sqrtwo}
    }    
    \foreach \j in {1}
    \foreach \i in {2, 3, 4, 5, 7, 8, 9}
    {
    \IdentityII{\i/\sqrtwo - 0.5/\sqrtwo}{\j/\sqrtwo}
    }
    \draw[thick, black] (-.5/\sqrtwo + 5.,1.5/\sqrtwo) -- (-.5/\sqrtwo + 5. + 0.25, 1.5/\sqrtwo + 0.25);    
    \draw[thick, black] (-.5/\sqrtwo + 5.,2.5/\sqrtwo) -- (-.5/\sqrtwo + 5. + 0.25, 2.5/\sqrtwo - 0.25);   
    \draw[thick, black] (-.5/\sqrtwo + 7.,0.5/\sqrtwo) -- (-.5/\sqrtwo + 7. - 0.25, 0.5/\sqrtwo - 0.25);
    \draw[thick, black] (-.5/\sqrtwo + 7.,-0.5/\sqrtwo) -- (-.5/\sqrtwo + 7. - 0.25, -0.5/\sqrtwo + 0.25);
    \node at (5.5/\sqrtwo, 2./\sqrtwo)[circle,fill,inner sep=0.4pt]{};
    \node at (5.5/\sqrtwo - 0.25, 2./\sqrtwo)[circle,fill,inner sep=0.4pt]{};
    \node at (5.5/\sqrtwo + 0.25, 2./\sqrtwo)[circle,fill,inner sep=0.4pt]{};
    \node at (5.5/\sqrtwo, 0./\sqrtwo)[circle,fill,inner sep=0.4pt]{};
    \node at (5.5/\sqrtwo - 0.25, 0./\sqrtwo)[circle,fill,inner sep=0.4pt]{};
    \node at (5.5/\sqrtwo + 0.25, 0./\sqrtwo)[circle,fill,inner sep=0.4pt]{};
    \node at (5.5/\sqrtwo, -1.1/\sqrtwo)[circle,fill,inner sep=0.4pt]{};
    \node at (5.5/\sqrtwo - 0.25, -1.1/\sqrtwo)[circle,fill,inner sep=0.4pt]{};
    \node at (5.5/\sqrtwo + 0.25, -1.1/\sqrtwo)[circle,fill,inner sep=0.4pt]{};
    \draw[thin, gray] (-.5/\sqrtwo,-1.1/\sqrtwo) -- (5./\sqrtwo,-1.1/\sqrtwo);
    \draw[thin, gray] ( 6./\sqrtwo,-1.1/\sqrtwo) -- (-.5/\sqrtwo + 9./\sqrtwo,-1.1/\sqrtwo);
    \foreach \i in {0,...,5}
    {
        \draw[thin, gray] (-.5/\sqrtwo + \i/\sqrtwo,-0.8) -- (-.5/\sqrtwo + \i/\sqrtwo,-1.1/\sqrtwo);
    }
    \foreach \i in {0,2, 4}
    {
        \Text[x=-0.5/\sqrtwo+\i/\sqrtwo,y=-1.5/\sqrtwo]{\tiny $\i$};
    }
    \foreach \i in {7,8, 9}
    {
        \draw[thin, gray] (-.5/\sqrtwo + \i/\sqrtwo,-0.8) -- (-.5/\sqrtwo + \i/\sqrtwo,-1.1/\sqrtwo);
    }
    \Text[x=-0.5/\sqrtwo+7/\sqrtwo,y=-1.6/\sqrtwo]{\tiny $L-2$};
    \Text[x=-0.5/\sqrtwo+9/\sqrtwo,y=-1.6/\sqrtwo]{\tiny $L$};
    \Text[x=5./\sqrtwo,y=-2.0/\sqrtwo]{\tiny $x$}
    \draw [thick, decorate,decoration={brace,amplitude=3pt,mirror}]
    (8.9, -0.5)--(8.9, 0.5) node[midway, yshift=0pt, xshift=12pt] {$\mathcal{S}_1$};
    \draw [thick, decorate,decoration={brace,amplitude=3pt,mirror}]
    (8.9, 1.5)--(8.9, 2.5) node[midway, yshift=0pt, xshift=12pt] {$\mathcal{S}_2$};
    \end{tikzpicture} 
     =: \mathcal{S}_2 V_{0,1} \mathcal{S}_1 
    \label{eq:Heisenberg_propagator}
\end{align}
which is shown here for odd $L$.
To this end we rewrite $\W$ in terms of the folded gate $V$ instead of $W$ and 
introduce the combined action of the swap gates $S_{x, x+1}$ acting on neighboring lattice sites $x$ and $x+1$ in the first (last) layer as
$\mathcal{S}_1$ ($\mathcal{S}_2$).
Additionally, we indicate the labels of the lattice sites above.
The matrix-product representation of $\W^t$ is obtained by contracting the inner legs corresponding to the swap gates to a compact form and by translating the swap gates which connect to in- and out-going legs, respectively, into suitable permutations of local lattice sites.
To make this more formal let us begin by introducing some notation.
We fix the size $L$ of the system and we denote the local Hilbert spaces obtained by the vectorization of local operators by $\K_x \simeq \mathds{C}^{q^2}$ for the lattice sites $x \in \{0, 1, \ldots, L\}$.
That is, $\W$ acts on $\K := \bigotimes_{x=0}^L\K_x \simeq \mathds{C}^{q^{2(L+1)}}$.
Moreover we define $\left[L\right]:=\{1, 2, \ldots, L\}$ and introduce the Hilbert space $\K_{\left[L\right]} := \bigotimes_{x \in \left[L\right]} \K_x$, i.e., $\K = \K_0 \otimes \K_{\left[L\right]}$.
By fixing an orthonormal basis $\{\ket{i}| i \in \{0, 1, \ldots, q^2-1\}\}$ of $\mathds{C}^{q^2}$ we obtain a product basis both in $\K$ and $\K_{\left[L\right]}$ whose elements we denote by 
$\ket{i_0 i_1\ldots i_L} := \ket{i_0} \otimes \ket{i_1} \otimes  \ldots \otimes \ket{i_L}$ and $\ket{i_1\ldots i_L}$, respectively. \\

The symmetric group $S_L$ on $L$ elements acts on $\K_{\left[L\right]}$ via the unitary representation $\Perm: S_L \to \text{U}(\K_{\left[L\right]})$
which permutes tensor factors.
More precisely, we identify elements of the product basis $\{\ket{i_1 \ldots i_L} | i_x \in \{0, \ldots q^2 -1 \}\}$ with maps $\mathbf{i}:\left[L\right] \to \{\ket{i} \in \{0, \ldots q^2 -1 \}\}$ via $\mathbf{i} \mapsto \ket{\mathbf{i}(1) \ldots \mathbf{i}(L)} = \ket{i_1 \ldots i_L}$. A left action of $S_L$ on the set of these maps is given by $\mathbf{i} \overset{\sigma}{\mapsto}\mathbf{i}\circ \sigma^{-1}$ 
for $\sigma \in S_L$. 
This translates to a left action of $S_L$ on the product basis which gives rise to the unitary operator $\Perm_\sigma$ defined by its action on basis vectors
\begin{align}
    \Perm_\sigma \ket{i_1 i_2 \ldots i_L} := 
    \ket{i_{\sigma^{-1}(1)} i_{\sigma^{-1}(2)} \ldots i_{\sigma^{-1}(L)}}
\end{align}
and linear extension. Note, that with this definition one has 
\begin{align}
\Perm_\rho \Perm_\sigma\ket{i_1\ldots i_L}=
\Perm_\rho \ket{i_{\sigma^{-1}(1)}\ldots i_{\sigma^{-1}(L)}} = 
\ket{i_{\sigma^{-1}\rho^{-1}(1)}\ldots i_{\sigma^{-1}\rho^{-1}(L)}} = 
\ket{i_{(\rho\sigma)^{-1}(1)}\ldots i_{(\rho\sigma)^{-1}(L)}} =
\Perm_{\rho\sigma}\ket{i_1\ldots i_L}
\end{align}
for any $\sigma, \rho \in S_L$.
We represent the unitary operator $\Perm_\sigma$ diagrammatically by a box as
\begin{align}
    \Perm_\sigma =
    \begin{tikzpicture}[baseline=-2, scale=0.6]
    \foreach  \x in {0, 1, 2, 4, 5}
    { 
        \draw[thick] (\x, -0.7) -- (\x, 0.7);
    }
    \draw[thick, fill=myblue, rounded corners=1pt] (-0.25, -0.3) rectangle  (0.25 + 5,0.3);
    \Text[x=3.,y=0.55]{$\dots$}
    \Text[x=3.,y=-1.1]{$\dots$}
    \Text[x=2.5,y=0]{$\sigma$}
    \draw[thin, gray] (0. , -1.1) -- (2.5, -1.1);
    \draw[thin, gray] (3.5 ,-1.1) -- (5, -1.1);
    \foreach \i in {0, 1, 2, 4, 5}
    {
        \draw[thin, gray] (\i ,-0.9) -- (\i ,-1.1 );
    }
    \Text[x=5.,y=-1.4]{\tiny $L$};
    \Text[x=4.,y=-1.4]{\tiny $L-1$};
    \Text[x=0.,y=-1.4]{\tiny $1$};
    \Text[x=1.,y=-1.4]{\tiny $2$};
    \Text[x=2.,y=-1.4]{\tiny $3$};
    \Text[x=3.5,y=-1.9]{\tiny $x$};
    \end{tikzpicture} 
\end{align}
for each permutation $\sigma \in S_L$. Let us make a remark about the diagrammatic technique used here: As $\Perm_\sigma$ maps simple tensors to simple tensors with only its tensor factors permuted the box in the above definition might be thought of $L$ wires each of which connects an in-going leg with an out-going leg. In a diagrammatic representation those wires might be arbitrarily continuously deformed without changing the operator as long as in-going and out-going legs are held fixed at the corresponding lattice site. \\

In preparation for our proof we consider specific permutations and study some of their relations in the following.
To this end it is convenient to adapt the following convention: For any integer $x \in \mathds{Z}$ we take the representative of $x\,\mod L$ in the set $\left[L\right]$, i.e., we identify $L\equiv 0$. 
Using this convention we define permutations $\sigma_\delta \in S_L$, indexed by $\delta \in \mathds{Z}$, by 
\begin{align}
\sigma_\delta(x) := \begin{cases}
2(x + \delta)\,\mod L & \text{if } (x+\delta)\,\mod L < \frac{L+1}{2}, \\
\left(-2(x + \delta) + 1\right)\, \mod L  & \text{if }(x + \delta)\,\mod L \geq \frac{L+1}{2}
\end{cases}
\end{align}
for $x \in \left[L\right]$.
A simple calculation shows that the inverse of $\sigma_\delta$ is given by
\begin{align}
\sigma^{-1}_\delta(x) := \begin{cases}
\left(\frac{x}{2} - \delta\right)\,\mod L & \text{if }x\text{ is even}, \\
\left(-\frac{x-1}{2} - \delta\right)\,\mod L & \text{if }x\text{ is odd}.
\end{cases}
\end{align}
By straightforward computations one moreover obtains the following properties. \\

\noindent \textbf{Lemma 1:} Given $\alpha, \delta \in \mathds{Z}$ one has
\begin{enumerate}[(i)]
    \item{$\sigma^{-1}_\alpha \sigma_{\delta + \alpha} = \eta_\delta \in S_L$, where
        $\eta_\delta$ denotes the periodic shift by $\delta$, i.e., 
        $\eta_\delta(x) := (x + \delta)\, \mod L$ for $x \in \left[L\right]$ and}
    \item{$\pi_\delta := \sigma_{\alpha} \sigma^{-1}_{\delta + \alpha}  \in S_L$ is given by
    \begin{align}
    \pi_\delta(x) =
    \begin{cases}
    (x - 2\delta)\,\mod L & \text{if } x \text{ is even and } \left(\frac{x}{2} - \delta \right)\,\mod L < \frac{L+1}{2}, \\
    (-x + 2\delta + 1)\,\mod L & \text{if } x \text{ is even and } \left(\frac{x}{2} - \delta \right)\,\mod L \geq \frac{L+1}{2}, \\
    (-x - 2\delta + 1)\,\mod L & \text{if } x \text{ is odd and } \left(-\frac{x-1}{2} - \delta \right)\,\mod L < \frac{L+1}{2}, \\
    (x + 2\delta)\,\mod L & \text{if } x \text{ is odd and } \left(-\frac{x-1}{2} - \delta \right)\,\mod L \geq \frac{L+1}{2}
    \end{cases} 
    \end{align}    
    for $x \in \left[L\right]$ and independent of $\alpha$. \\
}
\end{enumerate}

\noindent In case $\delta=1$ the second part of Lemma 1 reduces to 
    \begin{align}
\pi_1(x) =
\begin{cases}
x - 2 & \text{if } x \text{ is even and } x > 2 \\
1 & \text{if } x = 2 \\
L - 1 & \text{if } x \text{ is odd and } x = L \\ 
x + 2 & \text{if } x \text{ is odd and } x < 2
\end{cases}
\label{supp_eq:pi_delta}
\end{align}    
for $x \in \left[L\right]$. \\

As last preparatory step before formulating our claim we introduce operators $\R_\delta$ for $\delta \in \{0, 1, \ldots L\}$ acting on $\K$ in the form of two-layer matrix-product operators intertwined by their boundary conditions which are most conveniently defined by their respective diagrammatic representation
\begin{align}
    \R_\delta :=\begin{tikzpicture}[baseline=5, scale=0.6]
    \draw[thick] (7.5, 1.) arc (270:90: -0.5);
    \draw[thick] (-0.5, 0.) arc (90:180: 0.5);
    \draw[thick] (-1., 1.5) arc (180:270: 0.5);
    \foreach \y in {0, 1}
    \foreach \x in {0, 1, 3,4,5, 7}
    {
        \SwapR{\x}{\y}
    }
    \foreach \x in {4, 5, 7}
    \foreach \y in {1}
    {
        \transfermatrixgateR{\x}{\y}
    }
\draw[thin, gray] (-1. ,-1.1 ) -- (1.5, -1.1 );
\draw[thin, gray] (2.5 ,-1.1 ) -- (5.5, -1.1 );
\draw[thin, gray] (6.5 ,-1.1 ) -- (7., -1.1 );
\foreach \i in {-1, 0, 1, 3, 4, 5, 7}
{
    \draw[thin, gray] (\i ,-0.8) -- (\i ,-1.1 );
}
\Text[x=-1.,y=-1.4 ]{\tiny$0$};
\Text[x=1.,y=-1.4]{\tiny$2$};
\Text[x=3.,y=-1.4 ]{\tiny$ L-\delta$};
\Text[x=5.,y=-1.4 ]{\tiny$L-\delta + 2$};
\Text[x=7.,y=-1.4 ]{\tiny$L$};
\Text[x=6.5 ,y=-1.8 ]{\tiny$x$}
\foreach \x in {2, 6}
\foreach \y in {0, 1, -1.1}
{
\node at (\x, \y)[circle,fill,inner sep=0.4pt]{};
\node at (\x - 0.15, \y)[circle,fill,inner sep=0.4pt]{};
\node at (\x + 0.15, \y)[circle,fill,inner sep=0.4pt]{};
}
    \end{tikzpicture} \, .
\end{align}
Here all local tensors in the lower layer corresponding to lattice sites in $\left[L\right]$ as well as those in the upper layer corresponding to lattice sites in 
$\{1, 2,\ldots, L-\delta \}$ are given by swap gates.
In contrast, the local tensors at the remaining lattice sites in the upper layer are 
given by the folded gates $V$ with the orientation indicated by the mark in the 
respective top corner.
Note that $\R_0 = \mathds{1}_\K$.
One might also rewrite $\R_\delta$ as a quantum circuit with $2L$ layers and one nontrivial gate per layer as
\begin{align}
    \R_\delta  
    & = \quad \begin{tikzpicture}[baseline=102, scale=0.425]
 \foreach \x in {0, 1, 3, 4, 5, 7, 8}
 {
     \Swap{\x}{\x}
 }
  \foreach \x in { 0, 1, 3, 4, 5, 7, 8}
{
 \draw[thick, black] (\x + 0.5 ,-1. ) -- (\x + 0.5, \x - 0.5 );
 \draw[thick, black] (\x - 0.5 , \x + 0.5) -- (\x - 0.5, 17 - \x - 0.5 );
 \draw[thick, black] (\x + 0.5, 18 - \x - 0.5 ) -- (\x + 0.5, 18.);
}
\draw[thick, black] (- 0.5, 18. ) -- (- 0.5, 18. - 0.5);
\draw[thick, black] (- 0.5, -0.5 ) -- (- 0.5, -1.);
 \foreach \x in {0, 1, 3, 4, 5, 7, 8}
 {
 \Swap{8-\x}{\x + 9}
}
 \foreach \x in {4, 5, 7, 8}
 {
     \transfermatrixgateunrotated{\x}{8 - \x + 9}
 }%
\draw[thin, gray] (-0.5 ,-1.5 ) -- (2., -1.5);
\draw[thin, gray] (3,-1.5 ) -- (6., -1.5);
\draw[thin, gray] (7.,-1.5) -- (8.5, -1.5);
\foreach \i in {0,1, 2, 4, 5, 6, 8, 9}
{
\draw[thin, gray] (\i + -0.5 ,-1.2) -- (\i -0.5,-1.5 );
}
\Text[x=-0.5,y=-1.9 ]{\scriptsize $0$};
\Text[x=1.5,y=-1.9]{\scriptsize $2$};
\Text[x=4.5,y=-1.9 ]{\scriptsize $ L-\delta +1$};
\Text[x=8.5,y=-1.9 ]{\scriptsize $L$};
\Text[x=6.5 ,y=-2.5 ]{\scriptsize $x$}
\Text[x=2.5 ,y=-1.5 ]{\scriptsize $\ldots$}
\Text[x=6.5 ,y=-1.5 ]{\scriptsize $\ldots$}
\Text[x=2. ,y=2. ]{$\ldots$}
\Text[x=6.5 ,y=2. ]{$\ldots$}
\Text[x=2. ,y=15. ]{$\ldots$}
\Text[x=6.5 ,y=15. ]{$\ldots$}
\Text[x=1.5,y=8.5 ]{$\ldots$}
\Text[x=5.5 ,y=8.5]{$\ldots$}
\draw [thick, decorate,decoration={brace,amplitude=3pt,mirror}]
(9.2, -1.)--(9.2, 8.5) node[midway, yshift=0pt, xshift=12pt] {$\R_\delta^{\prime}$};
\draw [thick, decorate,decoration={brace,amplitude=3pt,mirror}]
(9.2, 13.5)--(9.2, 17.8) node[midway, yshift=0pt, xshift=12pt] {$\R_\delta^{\prime \prime}$};
\end{tikzpicture}  
 \label{supp_eq:R_delta2} 
 \!\!\!\!\!\!= \R_\delta^{\prime \prime}\left( 
 V_{L-\delta, L-\delta +1}\cdots V_{L-2,L} V_{L-1, L}\right) \R_\delta^{\prime}.
\end{align}
Here, we introduced $\R_\delta^{\prime}$ ($\R_\delta^{\prime \prime}$) which combines the action of the swap gates in the first $L$ (last $L-\delta$) layers.
The above rewriting of $\R_\delta$ additionally demonstrates its unitarity. \\

We are now able to state our main claim as \\

\noindent \textbf{Proposition:} For any non-negative integer $t$, written as $t = \tau L + \delta$ with unique non-negative integer $\tau$ and $\delta \in \{0, 1, \ldots, L-1\}$, one has
\begin{align}
    \W^t = \left(\mathds{1}_{\K_0} \otimes \Perm_{\sigma_\delta} \right)
    \R_\delta \R^\tau_L \left(\mathds{1}_{\K_0} \otimes \Perm_{\sigma_0}^{-1}  \right).
    \label{eq:MPO_for_propagator}
\end{align}

\noindent Intuitively, the first and the last factor describe the action of those swap gates 
which connect to out-going and in-going legs of a diagrammatic representation of $\W^t$ 
while the factor $\R_\delta\R^\tau_L$ combines the effect of the $t$ impurities at the 
systems boundary.
We prove the proposition by induction on $t$. \\

The case $t=0$ is trivial as $\tau=\delta=0$ and $\R_0 = 
\mathds{1}_{\K_{\left[L\right]}}$ yielding
\begin{align}
\left(\mathds{1}_{\K_0} \otimes \Perm_{\sigma_0} \right)
\R_0\left(\mathds{1}_{\K_0} \otimes \Perm_{\sigma_0}^{-1} \right) =
\left(\mathds{1}_{\K_0} \otimes \Perm_{\sigma_0}\Perm_{\sigma_0}^{-1} \right) = 
\mathds{1}_{\K} = \W^0.
\end{align}
However, we additionally need to consider the case $t=1$, i.e., $\tau=0$ and $\delta=1$, separately. 
It suffice to check that $\W \ket{i_0 \cdots i_L} = \left(\mathds{1}_{\K_0} \otimes \Perm_{\sigma_1}\right) \R_1 \left(\mathds{1}_{\K_0} \otimes \Perm_{\sigma_0}^{-1}\right) \ket{i_0 \cdots i_L}$ for any basis state $\ket{i_0 \cdots i_L}$.
We restrict ourselves to odd $L$, the case of even $L$ works analogously.
To this end, we use the factorization of $\W$ introduced in 
Eq.~\eqref{eq:Heisenberg_propagator}
and obtain
\begin{align}
\W \ket{i_0 i_1 \cdots i_L} & = \mathcal{S}_2 V_{0,1} \mathcal{S}_1 \ket{i_0 i_1 \cdots i_L} \\
& = \mathcal{S}_2V_{0, 1} \ket{i_1 i_0 i_3 i_2 \cdots i_L i_{L-1}} \\
& = \sum_{k,l} \mathcal{S}_2 V_{i_1i_0}^{kl}\ket{k\,l\,i_3 i_2 \cdots i_L i_{L-1}} \\
& = \sum_{k,l} V_{i_1i_0}^{kl}\ket{k\,i_3 l\,i_5 i_2 \cdots i_L i_{L-3} i_{L-1}}, 
\label{supp_eq:W_state}
\end{align}
where $V_{ij}^{kl} := \bra{kl}V\ket{ij}$ denotes the matrix elements of the folded gate $V$.
On the other hand we have
\begin{align}
\left(\mathds{1}_{\K_0} \otimes \Perm_{\sigma_1}\right) \R_1 \left(\mathds{1}_{\K_0} \otimes \Perm_{\sigma_0}^{-1}\right) \ket{i_0 \cdots i_L} & = 
\left(\mathds{1}_{\K_0} \otimes\Perm_{\sigma_1}\right)\R_1^{\prime \prime} 
V_{L-1, L}\R_1^\prime\left(\mathds{1}_{\K_0} \otimes \Perm_{\sigma_0}^{-1}\right) \ket{i_0 \cdots i_L} \\
& = \left(\mathds{1}_{\K_0} \otimes\Perm_{\sigma_1}\right)\R_1^{\prime \prime} 
V_{L-1, L}\R_1^\prime \ket{i_0 i_{\sigma_0(1)}\cdots i_{\sigma_0(L)}} \\
& = \left(\mathds{1}_{\K_0} \otimes\Perm_{\sigma_1}\right)\R_1^{\prime \prime} 
V_{L-1, L}\ket{i_{\sigma_0(1)}i_{\sigma_0(2)}\cdots i_{\sigma_0(L)} i_0} \\
& = \sum_{k,l} V_{i_{\sigma_0}(L) i_0}^{kl}
\left(\mathds{1}_{\K_0} \otimes\Perm_{\sigma_1}\right)\R_1^{\prime \prime}
\ket{i_{\sigma_0(1)}i_{\sigma_0(2)}\cdots k\,l}\\
& = \sum_{k,l} V_{i_1 i_0}^{kl}
\left(\mathds{1}_{\K_0} \otimes\Perm_{\sigma_1}\right)
\ket{k \,i_{\sigma_0(1)}i_{\sigma_0(2)}\cdots ,l}  \label{line1} \\
& = \sum_{k,l} V_{i_1 i_0}^{kl}
\ket{k \,i_{\sigma_0\sigma_1^{-1}(1)}l\,i_{\sigma_0\sigma_1^{-1}(3)}
    i_{\sigma_0\sigma_1^{-1}(4)} \cdots
    i_{\sigma_0\sigma_1^{-1}(L)}} \label{line2}\\
& = \sum_{k,l} V_{i_1 i_0}^{kl}
\ket{k \,i_{\pi_1(1)} l\,i_{\pi_1(3)}i_{\pi_1(4)} \cdots
i_{\pi_1(L)}}, \label{supp_eq:PR1P_state}
\end{align}
where in line~\eqref{line1} we used $\sigma_0(L)=1$ and in 
line~\eqref{line2} we used $\sigma_1(L)=2$. In the last line we inserted the definition of $\pi_1$ from Lemma 1 (ii). Using the lemma (and Eq.~\eqref{supp_eq:pi_delta}) we see that the expressions~\eqref{supp_eq:W_state}~and~\eqref{supp_eq:PR1P_state} coincide thereby proving our claim for $t=1$. \\

Now let us assume Eq.~\eqref{eq:MPO_for_propagator} is true for any $t^\prime \leq t$.
This implies the claim for $t+1$, corresponding to the step $\delta \to \delta + 1$, as we demonstrate in the following.
Note, that all preceding arguments work also for the borderline cases 
$\delta = 0$ and $\delta=L-1$.
In any case we have
\begin{align}
    \W^{t+1} = \W \W^t &  = \W \left(\mathds{1}_{\K_0} \otimes \Perm_{\sigma_\delta} \right)
    \R_\delta \R^\tau_L \left(\mathds{1}_{\K_0} \otimes \Perm_{\sigma_0}^{-1} \right) \\
    & = \left(\mathds{1}_{\K_0} \otimes \Perm_{\sigma_1} \right)
    \R_1 \left(\mathds{1}_{\K_0} \otimes \Perm_{\sigma_0}^{-1} \right)\left(\mathds{1}_{\K_0} \otimes \Perm_{\sigma_\delta} \right)
    \R_\delta \R^\tau_L \left(\mathds{1}_{\K_0} \otimes \Perm_{\sigma_0}^{-1} \right) \\
    & = \left(\mathds{1}_{\K_0} \otimes \Perm_{\sigma_1} \right)
    \R_1 \left(\mathds{1}_{\K_0} \otimes \Perm_{\sigma_0}^{-1}\Perm_{\sigma_\delta} \right)  \R_\delta \R^\tau_L \left(\mathds{1}_{\K_0} \otimes \Perm_{\sigma_0}^{-1} \right) \\
    & = \left(\mathds{1}_{\K_0} \otimes \Perm_{\sigma_1} \right)
    \R_1 \left(\mathds{1}_{\K_0} \otimes \Perm_{\eta_\delta} \right)  \R_\delta \R^\tau_L \left(\mathds{1}_{\K_0} \otimes \Perm_{\sigma_0}^{-1} \right),
    \label{Supp_eq:induction_step}
\end{align}
where we used the assumption for $\W^t$ ($\W^1$) in the first (second) line.
In the last line we used the fact, that $\Perm$ is a homomorphism as 
well as Lemma 1 (i). In order to show that the right hand side of 
Eq.~\eqref{Supp_eq:induction_step} equals $\left(\mathds{1}_{\K_0} \otimes \Perm_{\sigma_{\delta + 1}} \right)
\R_{\delta + 1} \R^\tau_L \left(\mathds{1}_{\K_0} \otimes \Perm_{\sigma_0}^{-1} \right)$
it suffices to show that 
\begin{align}
    \left(\mathds{1}_{\K_0} \otimes \Perm_{\sigma_{\delta + 1}} \right)
    \R_{\delta + 1} &  = \left(\mathds{1}_{\K_0} \otimes \Perm_{\sigma_1} \right)
    \R_1 \left(\mathds{1}_{\K_0} \otimes \Perm_{\eta_\delta} \right)  \R_\delta \\
    \Longleftrightarrow \qquad  \qquad \qquad \qquad \R_{\delta +1} & = \left(\mathds{1}_{\K_0} \otimes \Perm_{\sigma_{\delta + 1}}^{-1} \right)
   \left(\mathds{1}_{\K_0} \otimes \Perm_{\sigma_1} \right)
   \R_1 \left(\mathds{1}_{\K_0} \otimes \Perm_{\eta_\delta} \right)  \R_\delta \\
    & = \left(\mathds{1}_{\K_0} \otimes \Perm_{\eta_\delta}^{-1} \right)
    \R_1 \left(\mathds{1}_{\K_0} \otimes \Perm_{\eta_\delta} \right)  \R_\delta,
    \label{Supp_eq:induction_step2}
\end{align}
where in the last line we again used Lemma 1 (i). 
As $\Perm_{\eta_\delta}$ implements a periodic shift on the lattice sites in $\left[L\right]$ conjugation of $\R_1$ by $\left(\mathds{1}_{\K_0} \otimes \Perm_{\eta_\delta} \right)$ shifts the action of the only non-swap gate from lattice site $L$ to lattice site $L-\delta$, i.e., 
\begin{align}
    \left(\mathds{1}_{\K_0} \otimes \Perm_{\eta_\delta}^{-1} \right)
    \R_1 \left(\mathds{1}_{\K_0} \otimes \Perm_{\eta_\delta} \right) =
    \begin{tikzpicture}[baseline=-2, scale=0.6]
    \foreach \y in {-1, 2}
    \foreach  \x in {0, 1, 3, 4, 5, 7}
    { 
        \draw[thick] (\x, \y -0.7) -- (\x, \y + 0.7);
    }
    \draw[thick, fill=myblue, rounded corners=1pt] (-0.25, -1. -0.3) rectangle  (0.25 + 7,-1. + 0.3);
    \Text[x=3.5,y=-1.]{$\eta_\delta$}
    \draw[thick, fill=myblue, rounded corners=1pt] (-0.25, 2. -0.3) rectangle  (0.25 + 7,2. + 0.3);
    \Text[x=3.5,y=2]{$\eta_{-\delta}$}
    \draw[thick] (-1, -1.7) -- (-1., -0.5);
    \draw[thick] (-1, 2.7) -- (-1., 1.5);
    \draw[thick] (7.5, 1.) arc (270:90: -0.5);
    \draw[thick] (-0.5, 0.) arc (90:180: 0.5);
    \draw[thick] (-1., 1.5) arc (180:270: 0.5);
    \foreach \y in {0, 1}
    \foreach \x in {0, 1, 3,4,5, 7}
    {
        \SwapR{\x}{\y}
    }
    \foreach \x in {7}
    \foreach \y in {1}
    {
        \transfermatrixgateR{\x}{\y}
    }
    \draw[thin, gray] (-1. ,-2.1 ) -- (1.5, -2.1 );
    \draw[thin, gray] (2.5 ,-2.1 ) -- (5.5, -2.1 );
    \draw[thin, gray] (6.5 ,-2.1 ) -- (7., -2.1 );
    \foreach \i in {-1, 0, 1, 3, 4, 5, 7}
    {
        \draw[thin, gray] (\i ,-1.9) -- (\i ,-2.1 );
    }
    \Text[x=-1.,y=-2.4 ]{\tiny$0$};
    \Text[x=1.,y=-2.4]{\tiny$2$};
    \Text[x=4.,y=-2.4 ]{\tiny$ L-\delta$};
    \Text[x=7.,y=-2.4 ]{\tiny$L$};
    \Text[x=6 ,y=-2.7 ]{\tiny$x$}
    \foreach \x in {2, 6}
    \foreach \y in {0, 1, -2.1, 2.6}
    {
        \node at (\x, \y)[circle,fill,inner sep=0.4pt]{};
        \node at (\x - 0.15, \y)[circle,fill,inner sep=0.4pt]{};
        \node at (\x + 0.15, \y)[circle,fill,inner sep=0.4pt]{};
    }
    \end{tikzpicture}\, =
    \begin{tikzpicture}[baseline=5, scale=0.6]
    \draw[thick] (7.5, 1.) arc (270:90: -0.5);
    \draw[thick] (-0.5, 0.) arc (90:180: 0.5);
    \draw[thick] (-1., 1.5) arc (180:270: 0.5);
    \foreach \y in {0, 1}
    \foreach \x in {0, 1, 3,4,5, 7}
    {
        \SwapR{\x}{\y}
    }
    \foreach \x in {4}
    \foreach \y in {1}
    {
        \transfermatrixgateR{\x}{\y}
    }
    \draw[thin, gray] (-1. ,-1.1 ) -- (1.5, -1.1 );
    \draw[thin, gray] (2.5 ,-1.1 ) -- (5.5, -1.1 );
    \draw[thin, gray] (6.5 ,-1.1 ) -- (7., -1.1 );
    \foreach \i in {-1, 0, 1, 3, 4, 5, 7}
    {
        \draw[thin, gray] (\i ,-0.8) -- (\i ,-1.1 );
    }
    \Text[x=-1.,y=-1.4 ]{\tiny$0$};
    \Text[x=1.,y=-1.4]{\tiny$2$};
    \Text[x=4.,y=-1.4 ]{\tiny$ L-\delta$};
    \Text[x=7.,y=-1.4 ]{\tiny$L$};
    \Text[x=6 ,y=-1.7 ]{\tiny$x$}
    \foreach \x in {2, 6}
    \foreach \y in {0, 1, -1.1}
    {
        \node at (\x, \y)[circle,fill,inner sep=0.4pt]{};
        \node at (\x - 0.15, \y)[circle,fill,inner sep=0.4pt]{};
        \node at (\x + 0.15, \y)[circle,fill,inner sep=0.4pt]{};
    }
    \end{tikzpicture}\, ,
\end{align}
which might also be checked on the basis vectors by using the representation~\eqref{supp_eq:R_delta2}.
Here we used $\eta_{-\delta} = \eta_{\delta}^{-1}$ in the diagrammatic representation.
Using the above equation, the right hand side of Eq.~\eqref{Supp_eq:induction_step2} can be evaluated diagrammatically (or by again using a similar representation as in Eq~\eqref{supp_eq:R_delta2}) as 
\begin{align}
\left(\mathds{1}_{\K_0} \otimes \Perm_{\eta_\delta}^{-1} \right)
\R_1 \left(\mathds{1}_{\K_0} \otimes \Perm_{\eta_\delta} \right)\R_\delta & = 
    \begin{tikzpicture}[baseline=-11, scale=0.6]
\draw[thick] (7.5, 1.) arc (270:90: -0.5);
\draw[thick] (-0.5, 0.) arc (90:180: 0.5);
\draw[thick] (-1., 1.5) arc (180:270: 0.5);
\foreach \y in {0, 1}
\foreach \x in {0, 1, 3,4,5, 7}
{
    \SwapR{\x}{\y}
}
\foreach \x in {4}
\foreach \y in {1}
{
    \transfermatrixgateR{\x}{\y}
}
\draw[thin, gray] (-1. ,-3.1 ) -- (1.5, -3.1 );
\draw[thin, gray] (2.5 ,-3.1 ) -- (5.5, -3.1 );
\draw[thin, gray] (6.5 ,-3.1 ) -- (7., -3.1 );
\foreach \i in {-1, 0, 1, 3, 4, 5, 7}
{
    \draw[thin, gray] (\i ,-2.8) -- (\i ,-3.1 );
}
\Text[x=-1.,y=-3.4 ]{\tiny$0$};
\Text[x=1.,y=-3.4]{\tiny$2$};
\Text[x=4.,y=-3.4 ]{\tiny$ L-\delta$};
\Text[x=7.,y=-3.4 ]{\tiny$L$};
\Text[x=6 ,y=-3.7 ]{\tiny$x$}
\foreach \x in {2, 6}
\foreach \y in {0, 1, -1, -2, -3.1}
{
    \node at (\x, \y)[circle,fill,inner sep=0.4pt]{};
    \node at (\x - 0.15, \y)[circle,fill,inner sep=0.4pt]{};
    \node at (\x + 0.15, \y)[circle,fill,inner sep=0.4pt]{};
}
\draw[thick] (7.5, -1.) arc (270:90: -0.5);
\draw[thick] (-0.5, -2.) arc (90:180: 0.5);
\draw[thick] (-1., -0.5) arc (180:270: 0.5);
\foreach \y in {-1, -2}
\foreach \x in {0, 1, 3,4,5, 7}
{
    \SwapR{\x}{\y}
}
\foreach \x in {5, 7}
\foreach \y in {-1}
{
    \transfermatrixgateR{\x}{\y}
}
\end{tikzpicture}
\\ 
& = \begin{tikzpicture}[baseline=5, scale=0.6]
\draw[thick] (7.5, 1.) arc (270:90: -0.5);
\draw[thick] (-0.5, 0.) arc (90:180: 0.5);
\draw[thick] (-1., 1.5) arc (180:270: 0.5);
\foreach \y in {0, 1}
\foreach \x in {0, 1, 3,4,5, 7}
{
    \SwapR{\x}{\y}
}
\foreach \x in {4, 5, 7}
\foreach \y in {1}
{
    \transfermatrixgateR{\x}{\y}
}
\draw[thin, gray] (-1. ,-1.1 ) -- (1.5, -1.1 );
\draw[thin, gray] (2.5 ,-1.1 ) -- (5.5, -1.1 );
\draw[thin, gray] (6.5 ,-1.1 ) -- (7., -1.1 );
\foreach \i in {-1, 0, 1, 3, 4, 5, 7}
{
    \draw[thin, gray] (\i ,-0.8) -- (\i ,-1.1 );
}
\Text[x=-1.,y=-1.4 ]{\tiny$0$};
\Text[x=1.,y=-1.4]{\tiny$2$};
\Text[x=4.,y=-1.4 ]{\tiny$ L-\delta$};
\Text[x=7.,y=-1.4 ]{\tiny$L$};
\Text[x=6 ,y=-1.7 ]{\tiny$x$}
\foreach \x in {2, 6}
\foreach \y in {0, 1, -1.1}
{
    \node at (\x, \y)[circle,fill,inner sep=0.4pt]{};
    \node at (\x - 0.15, \y)[circle,fill,inner sep=0.4pt]{};
    \node at (\x + 0.15, \y)[circle,fill,inner sep=0.4pt]{};
}
\end{tikzpicture} \\
& = \R_{\delta +1}.
\end{align}
This proves the claim for $t+1$ and hence completes the proof of the proposition. As a final remark we note that the above construction might be repeated for the unfolded circuit $\U$ in a similar fashion as no argument depends on the origin of the local Hilbert spaces $\K_x$ from the vectorization of operators. 
However, one needs to take the placement of the impurity interaction in the second rather than the first layer of the circuit into account.\\

We now come back to the computation of correlation functions $C_{ab}(t)$ for local operators $a_0$ and $b_0$ acting nontrivially only on the first site, i.e., site $0$. In vectorized form $a_0$ reads
\begin{align}
    \ket{a_0} = \ket{a}\otimes\ket{\circ}^{\otimes L} 
\end{align}
and is obviously an eigenvector of $\mathds{1}_{\K_0} \otimes \Perm_\sigma$ with eigenvalue $1$ for any permutation $\sigma \in S_L$.
The same of course holds for $b_0$.
Hence the correlation function at time $t=\tau L + \delta$ is given by
\begin{align}
    C_{ab}(t) & = \bra{b_0}\W^t\ket{a_0} = \bra{b_0}\R_\delta \R_L^\tau \ket{a_0} \\
    & =     \begin{tikzpicture}[baseline=-53, scale=0.6]
    \foreach  \x in {-1, 0, 1, 3, 4, 5, 7}
    {
    \draw[thick, black] (\x ,-7.5 ) -- (\x, -7.7 );
    \draw[thick, black] (\x ,1.5 ) -- (\x, 1.7 );
    \IdState{\x}{1.7}
    \IdState{\x}{-7.7}
    }
    \localOperator{-1}{1.7}
    \localOperator{-1}{-7.7}
    \foreach  \y in {0, -2, -5, -7}
    {
    \draw[thick] (7.5, 1. + \y) arc (270:90: -0.5);
    \draw[thick] (-0.5, 0. + \y) arc (90:180: 0.5);
    \draw[thick] (-1., 1.5 + \y) arc (180:270: 0.5);
    }
    \foreach \y in {0, 1}
    \foreach \x in {0, 1, 3,4,5, 7}
    {
        \SwapR{\x}{\y}
    }
    \foreach \x in {4,5, 7}
    \foreach \y in {1}
    {
        \transfermatrixgateR{\x}{\y}
    }
    \foreach \y in {-1, -2, -4, -5, -6, -7}
    \foreach \x in {0, 1, 3,4,5, 7}
    {
        \SwapR{\x}{\y}
    }
    \foreach \x in {0, 1, 3,4,5, 7}
    \foreach \y in {-1, -4, -6}
    {
        \transfermatrixgateR{\x}{\y}
    }
    \draw[thin, gray] (-1. ,-8.5 ) -- (1.5, -8.5 );
    \draw[thin, gray] (2.5 ,-8.5 ) -- (5.5, -8.5 );
    \draw[thin, gray] (6.5 ,-8.5 ) -- (7., -8.5 );
    \foreach \i in {-1, 0, 1, 3, 4, 5, 7}
    {
        \draw[thin, gray] (\i ,-8.2) -- (\i ,-8.5 );
    }
    \Text[x=-1.,y=-8.8]{\tiny$0$};
    \Text[x=1.,y=-8.8]{\tiny$2$};
    \Text[x=4.,y=-8.8 ]{\tiny$ L-\delta + 1$};
    \Text[x=7.,y=-8.8 ]{\tiny$L$};
    \Text[x=6 ,y=-9.1 ]{\tiny$x$}
    \foreach \x in {2, 6}
    \foreach \y in {0, 1, -1, -2, -4, -5, -6, -7, -8.5}
    {
        \node at (\x, \y)[circle,fill,inner sep=0.4pt]{};
        \node at (\x - 0.15, \y)[circle,fill,inner sep=0.4pt]{};
        \node at (\x + 0.15, \y)[circle,fill,inner sep=0.4pt]{};
    }
    \foreach \x in {-1, 0, 1, 3, 4, 5, 7}
    \foreach \y in {-3}
    {
        \node at (\x, \y)[circle,fill,inner sep=0.4pt]{};
        \node at (\x, \y - 0.15)[circle,fill,inner sep=0.4pt]{};
        \node at (\x, \y + 0.15)[circle,fill,inner sep=0.4pt]{};
    }
    \Text[x=-1.4,y=-7.7 ]{$a$};
    \Text[x=-1.4,y=1.8 ]{$b$};
    \draw [thick, stealth-stealth](8.5, - 7.) -- (8.5, -1.);
    \Text[x=8.8,y=-4.]{$\tau$}
    \end{tikzpicture}
     =   \, \, \,  \begin{tikzpicture}[baseline=-35, scale=0.6]
    \foreach  \x in {0, 1, 3, 4, 5, 7, 8}
    {
    \draw[thick, black] (\x ,-6.5 ) -- (\x, -6.7 );
    \draw[thick, black] (\x ,2.5 ) -- (\x, 2.7 );
    \IdState{\x}{2.7}
    \IdState{\x}{-6.7}
}
    \localOperator{8.}{2.7}
    \localOperator{8.}{-6.7}
    \foreach  \y in {0, -2, -5}
    {
        \draw[thick] (7.5, 1. + \y) arc (270:90: -0.5);
        \draw[thick] (-0.5, 0. + \y) arc (90:180: 0.5);
        \draw[thick] (-1., 1.5 + \y) arc (180:270: 0.5);
    }
    \draw[thick] (8., 2.5) arc (180:90: -0.5);
    \draw[thick] (-1., 1.5) arc (180:90: 0.5);
    \draw[thick] (7.5, -6.) arc (270:180: -0.5);
    \draw[thick] (-1., -5.5) arc (180:270: 0.5);
    \foreach \y in {2, 1, 0, -1, -2, -4, -5, -6,}
    \foreach \x in {0, 1, 3,4,5, 7}
    {
        \SwapR{\x}{\y}
    }
    \foreach \x in {4,5, 7}
    \foreach \y in {1}
    {
        \transfermatrixgateR{\x}{\y}
    }
    \foreach \x in {0, 1, 3,4,5, 7}
    \foreach \y in {-1, -4, -6}
    {
        \transfermatrixgateR{\x}{\y}
    }
    \foreach \x in {2, 6}
    \foreach \y in {1, 0, 1, -1, -2, -4, -5, -6}
    {
        \node at (\x, \y)[circle,fill,inner sep=0.4pt]{};
        \node at (\x - 0.15, \y)[circle,fill,inner sep=0.4pt]{};
        \node at (\x + 0.15, \y)[circle,fill,inner sep=0.4pt]{};
    }
    \foreach \x in {-1, 0, 1, 3, 4, 5, 7}
    \foreach \y in {-3}
    {
        \node at (\x, \y)[circle,fill,inner sep=0.4pt]{};
        \node at (\x, \y - 0.15)[circle,fill,inner sep=0.4pt]{};
        \node at (\x, \y + 0.15)[circle,fill,inner sep=0.4pt]{};
    }
    \Text[x=8.4,y=-6.7 ]{$a$};
    \Text[x=8.4,y=2.8 ]{$b$};
    \end{tikzpicture} ,
\end{align}
where in the last step we contracted the first layer of swap gates and added the top layer minding that the internal wires corresponding to the swap gates can undergo arbitrary continuous deformations without changing the value of $C_{ab}(t)$.
The above diagrammatic representation is, however, just 
a ninety-degree counterclockwise rotated version of Eq.~\eqref{eq:correlations_via_transfer_matrix_diagramm}.
This concludes the proof of the equality claimed in Eqs.~\eqref{eq:correlations_via_transfer_matrix}~and~\eqref{eq:correlations_via_transfer_matrix_diagramm}.

\section{Spectral Properties of Transfer Matrices}
\label{S-IV}

In this section we present some numerical results on the spectral properties
of transfer matrices $\T_\tau$ for both T-dual and generic impurity interactions as well as various local Hilbert space dimensions (of the original spatial lattice) $q \in \{2, 3, 4\}$.
As $\T_\tau$ is real in the sense that it maps the vectorization of Hermitian operators to the vectorization of Hermitian operators its eigenvalues are either real or come in
complex conjugate pairs.
We depict the respective spectral densities of nontrivial eigenvalues of the transfer matrices for the ensembles introduced in the previous sections
in Fig.~\ref{fig:spectral_statistics}.
In any case we find the distribution to be a superposition of the distribution of real
eigenvalues and the distribution of complex conjugates pairs of eigenvalues.
The former is clearest seen for the T-dual case, where it is supported on the whole interval 
$[-1, 1]$, and is less pronounced and of strictly smaller support in the generic case.
In the T-dual case the distribution of complex eigenvalues has highest
weight on a ring with mean radius
approximately $1/2$ for both $q=3$ and $q=4$, with $q=2$ again showing different behavior.
In contrast, for generic impurities the distribution is supported mainly within the disk of
radius $1/q$.
From our numerical experiments we find these distributions to be stable upon varying $\tau$ (not shown).

Therefore one might hope, that also the leading nontrivial eigenvalue $\lambda_0=\lambda_0(\tau)$ does not grow when increasing $\tau$ at least when $\tau$ is
sufficiently large.
In Ref~\cite{KosBerPro2021} it is demonstrated that this happens typically for $q=2$.
As $|\lambda_0|$ is a monotonically non-decreasing function of $\tau$ which is bounded by $1$ the leading nontrivial eigenvalue converges as $\tau  \to \infty$.
In order to estimate whether this limit is smaller than $1$ and hence whether there is a spectral gap $\Delta_0 > 0$ we compute the leading
nontrivial eigenvalues for random realizations of the impurity interaction drawn from the ensembles introduced in the main text for various $\tau$. 
From this we estimate the probability 
$p=p(|\lambda_0(\tau + 1)| > |\lambda_0(\tau)|)$ for the leading nontrivial eigenvalue to grow when increasing $\tau$.
In order to efficiently compute the leading nontrivial eigenvalue we use the following numerical protocol:
We project a random initial vector, which is orthogonal to the trivial eigenvector, to the Jordan subspace (eigenspaces in the T-dual case) corresponding to the leading nontrivial eigenvalues.
This is achieved by applying $\T_\tau^k$ to the initial state for sufficiently large $k$ using the matrix product structure of $\T_\tau$.
Subsequently we construct the Krylov subspace containing the leading nontrivial eigenvalue's Jordan subspace via Arnoldi iteration applied to the projected initial vector taking also the possibility of a complex conjugate pair of leading nontrivial eigenvalues into account.
This allows to estimate the leading eigenvalue of $\T_\tau$ by diagonalizing $\T_\tau$ within the Krylov subspace.
However, as a consequence of $\T_\tau$ in general 
exhibiting a nontrivial Jordan structure, the accuracy of the obtained eigenvalues is relatively low. 
Nevertheless, this procedure allows to estimate 
$p(|\lambda_0(\tau + 1)| > |\lambda_0(\tau)|)$ as it is shown in
Fig.~\ref{fig:leading_eigenvalues}.
The probability decreases fast with $\tau$ and we find no increase in the leading eigenvalue at the largest accessible system sizes, except for T-dual gates at $\tau=4$.
More precisely for $q=2$ ($q=3$) we compute the leading eigenvalue up to a maximum value of $\tau=11$ ($8$) and find no increase in the leading eigenvalue for $\tau>5$.
For $q=4$ we compute the leading eigenvalue only up to a maximum value of $\tau=5$ and therefore Fig.~\ref{fig:leading_eigenvalues} does not include data for
$p(|\lambda_0(\tau + 1)| > |\lambda_0(\tau)|)$ for $\tau>4$ if $q=4$.
We find no increase in the leading eigenvalue already for $\tau>2$ for generic impurities and $q=4$.
In contrast, for T-dual impurities we find a nonzero probability for the leading eigenvalue to increase even at the largest accessible $\tau$.
Nevertheless $p(|\lambda_0(\tau + 1)| > |\lambda_0(\tau)|)$ is decreasing with $\tau$ for $\tau>2$ and we can hope for it to approach zero as well for larger $\tau$ than what we can access numerically.
Based on the above observations we conjecture, that generically the leading eigenvalue will become stationary for sufficiently large $\tau$ implying a nonzero spectral gap $\Delta_0$ and hence also a nonzero spectral gap for the leading relevant eigenvalue $\Delta_1 > 0$ in the T-dual case.

\begin{figure}[]
    \includegraphics{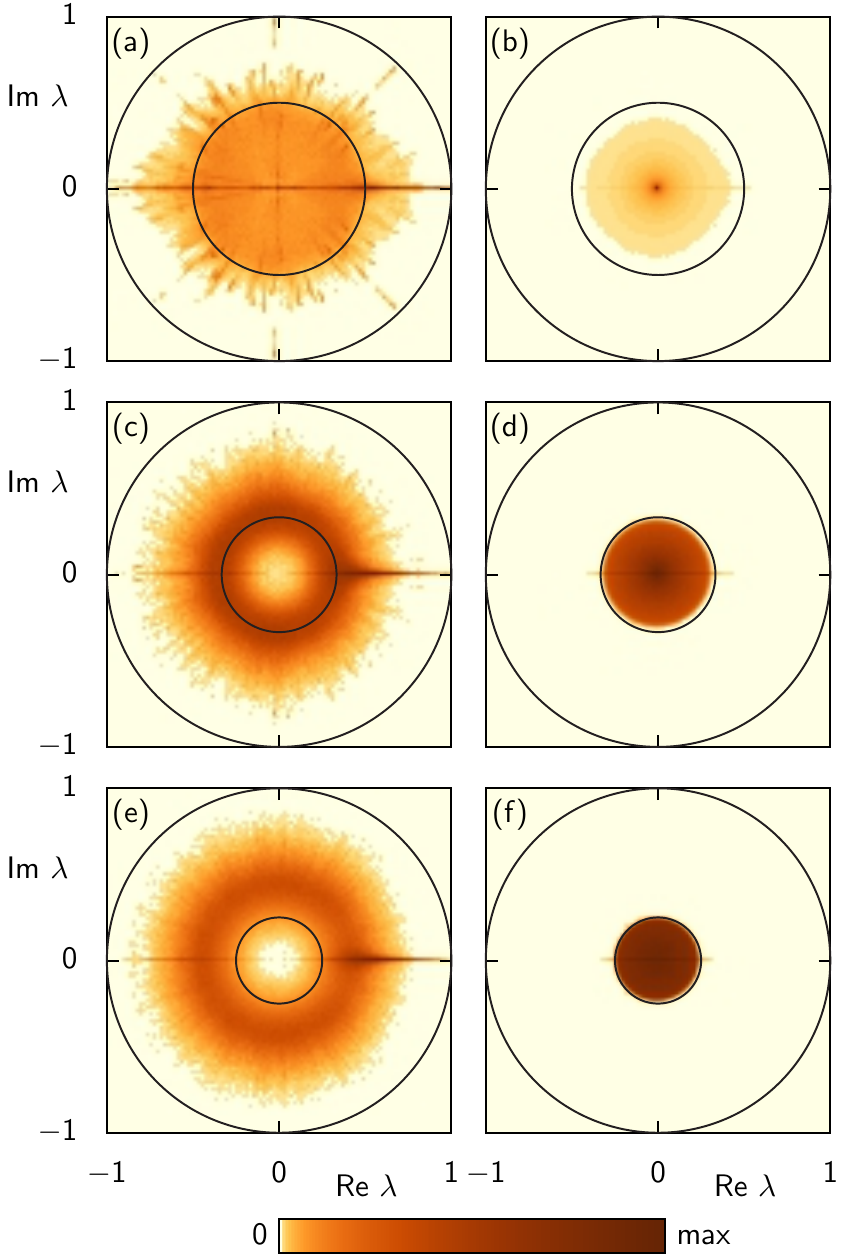}
    \caption{Histogram of the nontrivial eigenvalues for $q=2$, $\tau=7$ (a, b), 
        $q=3$, $\tau=4$ (c, d), and
        $q=4$, $\tau=3$ (e, f) for the T-dual case (a, c, e) as well as the generic
        case (b, d, f) for 500 realizations of the circuit for each case.
        The black circles have radii $1$ and $1/q$, respectively.
        Each eigenvalue is weighted by the degree of its degeneracy in the T-dual case and
        by the dimension of the corresponding Jordan block in the generic case.}
    \label{fig:spectral_statistics}
\end{figure}


\section{Eigenstates of transfer matrices in the T-dual case} 
\label{S-V}

In this section we motivate the notion of relevant eigenvectors (eigenoperators) in the case of T-dual impurities and derive the asymptotics of correlation functions given by Eq.~\eqref{eq:correlations_asymptotics}.
To this end we first construct the spectral decomposition
\begin{align}
    \T_\tau = \sum_\lambda \lambda \Proj_{\lambda, \tau}.
    \label{Supp_ex:spectral_decomposition}
\end{align}
More precisely, we demonstrate that the projections $\Proj_{\lambda, \tau}$ onto the eigenspaces corresponding to nontrivial eigenvalues $\lambda \in \text{spec}(\T_\tau)$ are given by Eq.~\eqref{eq:proj_dual_unitary}.
To this end, we show that the right eigenvectors $\ket{r_\lambda, s}$, Eq.~\eqref{eq:eigenvectors}, and the corresponding left eigenvectors $\bra{l_\lambda, s}$ are biorthogonal.
We initially restrict to the nontrivial 
eigenvalues
and extend the notation from the main text by adding a
subscript $\tau$ in the bra-ket notation, e.g., we write $\ket{\phi}_\tau, \ket{\psi}_\tau 
\in \mathds{C}^{q^{2\tau}}$ as well as $_\tau\!\bra{\psi}$ for the respective bra-vector in order to indicate the number $\tau$ of lattice sites.
We moreover denote scalar products of such vectors by $\braket{\phi | \psi}_\tau$ and write $\ket{\phi}\!\bra{\psi}_\tau$ for rank-one operators.
Additionally, we denote the right (left) eigenvectors for 
eigenvalue $\lambda$ of $\T_\tau$ with $\tau_\lambda = \tau$ by $\ket{r_\lambda}_{\tau_\lambda}$ ($_{\tau_\lambda} \!\!\bra{l_\lambda}$) and use 
the graphical representation
\begin{align}
\ket{r_\lambda}_{\tau_\lambda} = 
    \begin{tikzpicture}[baseline=0, scale=0.6]
    \foreach  \x in {0, 1, 2, 4, 5}
    { 
        \draw[thick] (\x, 0.) -- (\x, 0.8);
    }
\draw[thick, fill=myorange2, rounded corners=1pt] (-0.25, -0.3) rectangle  (0.25 + 5,0.3);
\Text[x=3.,y=0.55]{$\dots$}
\Text[x=2.5,y=0]{$\lambda$}
    \end{tikzpicture}\quad \text{and} \quad
_{\tau_\lambda}\!\!\bra{l_\lambda} = 
    \begin{tikzpicture}[baseline=-7, scale=0.6]
    \foreach  \x in {0, 1, 2, 4, 5}
    { 
        \draw[thick] (\x, 0.) -- (\x, -0.8);
    }
\draw[thick, fill=myorange2, rounded corners=1pt] (-0.25, -0.3) rectangle  (0.25 + 5,0.3);
\Text[x=3.,y=-0.55]{$\dots$}
\Text[x=2.5,y=0]{$\lambda$}
    \end{tikzpicture} \, ,
\label{eq:eigenvector_diagramm}
\end{align}
where the number of out-going legs and, respectively, in-going legs is $\tau_\lambda$.
These eigenvectors obey 
\begin{align}
\ket{r_\lambda, +}_{\tau_\lambda - 1} := 
\begin{tikzpicture}[baseline=0, scale=0.6]
\foreach  \x in {0, 1, 2, 4, 5}
{ 
    \draw[thick] (\x, 0.) -- (\x, 0.8);
}
\draw[thick, fill=myorange2, rounded corners=1pt] (-0.25, -0.3) rectangle  (0.25 + 5,0.3);
\Text[x=3.,y=0.55]{$\dots$}
\Text[x=2.5,y=0]{$\lambda$}
\IdState{5}{0.9}
\end{tikzpicture}\, = 0 = \
\begin{tikzpicture}[baseline=0, scale=0.6]
\foreach  \x in {0, 1, 2, 4, 5}
{ 
    \draw[thick] (\x, 0.) -- (\x, 0.8);
}
\draw[thick, fill=myorange2, rounded corners=1pt] (-0.25, -0.3) rectangle  (0.25 + 5,0.3);
\Text[x=3,y=0.55]{$\dots$}
\Text[x=2.5,y=0]{$\lambda$}
\IdState{0.}{0.9}
\end{tikzpicture}\, =: 
\ket{r_\lambda, -}_{\tau_\lambda - 1},
\label{eq:reduced_eigenvectors1} \\
_{\tau_\lambda - 1}\!\!\bra{l_\lambda, +} := 
\begin{tikzpicture}[baseline=-7, scale=0.6]
\foreach  \x in {0, 1, 2, 4, 5}
{ 
    \draw[thick] (\x, 0.) -- (\x, -0.8);
}
\draw[thick, fill=myorange2, rounded corners=1pt] (-0.25, -0.3) rectangle  (0.25 + 5,0.3);
\Text[x=3.,y=-0.55]{$\dots$}
\Text[x=2.5,y=0]{$\lambda$}
\IdState{5}{-0.9}
\end{tikzpicture}\, = 0 = \,
\begin{tikzpicture}[baseline=-7, scale=0.6]
\foreach  \x in {0, 1, 2, 4, 5}
{ 
    \draw[thick] (\x, 0.) -- (\x, -0.8);
}
\draw[thick, fill=myorange2, rounded corners=1pt] (-0.25, -0.3) rectangle  (0.25 + 5,0.3);
\Text[x=3,y=-0.55]{$\dots$}
\Text[x=2.5,y=0]{$\lambda$}
\IdState{0.}{-0.9}
\end{tikzpicture}\, =: 
\, _{\tau_\lambda - 1}\!\!\bra{l_\lambda, -}.
\label{eq:reduced_eigenvectors2}
\end{align}
Note, that contraction with the states $\ket{\circ}$ at a given subset of lattice sites 
corresponds to taking the partial trace with respect to the corresponding subsystem 
upon reversing the operator-state mapping.
Moreover, this demonstrates that the corresponding operators have full support on the $\tau$ lattice.

Equations~\eqref{eq:reduced_eigenvectors1}~and~~\eqref{eq:reduced_eigenvectors2} can be obtained using unitality, Eq.~\eqref{eq:unitality}, and dual unitality, 
Eq.~\eqref{eq:dual_unitality}, of the gate $V$.
For instance we find
\begin{align}
    \T_{\tau_\lambda - 1} \ket{r_\lambda, +}_{\tau_\lambda - 1} =
    \begin{tikzpicture}[baseline=0, scale=0.6]
    \foreach  \x in {0, 1, 2, 4, 5}
    { 
        \draw[thick] (\x, 0.) -- (\x, 0.8);
    }
    \draw[thick, fill=myorange2, rounded corners=1pt] (-0.25, -0.3) rectangle  (0.25 + 5,0.3);
    \Text[x=3.,y=1.]{$\dots$}
    \Text[x=2.5,y=0]{$\lambda$}
    \IdState{5}{0.7}
    \foreach \x in {0,1, 2, 4}
    \foreach \y in {1}
    {
        \transfermatrixgate{\x}{\y}
    }
    \foreach \y in {1}
    {
        \IdState{-0.6}{\y}
        \IdState{4.6}{\y}
    }
    \end{tikzpicture}\, = 
        \begin{tikzpicture}[baseline=0, scale=0.6]
    \foreach  \x in {0, 1, 2, 4, 5}
    { 
        \draw[thick] (\x, 0.) -- (\x, 0.8);
    }
    \draw[thick, fill=myorange2, rounded corners=1pt] (-0.25, -0.3) rectangle  (0.25 + 5,0.3);
    \Text[x=3.,y=1.]{$\dots$}
    \Text[x=2.5,y=0]{$\lambda$}
    \IdState{5}{1.6}
    \foreach \x in {0,1, 2, 4, 5}
    \foreach \y in {1}
    {
        \transfermatrixgate{\x}{\y}
    }
    \foreach \y in {1}
    {
        \IdState{-0.6}{\y}
        \IdState{5.6}{\y}
    }
    \end{tikzpicture}\, = \lambda \ket{r_\lambda, +}_{\tau_\lambda - 1},
 \end{align}
 where the second equality holds due to unitality of $V$ (in backward time direction) and the third equality reflects the fact that $\T_{\tau_\lambda}\ket{r_\lambda}_{\tau_\lambda}
 = \lambda \ket{r_\lambda}_{\tau_\lambda}$.
 Hence if $\ket{r_\lambda, +}_{\tau_\lambda - 1}$ was nonzero we would have  $\lambda \in \text{spec}(\T_{\tau_\lambda - 1})$ in contrast to $\tau_\lambda$ being minimal.
 This argument might be repeated for the remaining cases yielding all of Eqs.~\eqref{eq:reduced_eigenvectors1}~and~\eqref{eq:reduced_eigenvectors2}. \\
 
From this we prove the biorthogonality of the left and right eigenvectors of $\T_\tau$ for fixed $\tau$.
 For $\lambda, \sigma \in \text{spec}(\T_\tau)$ with
 $\tau_\lambda \neq \tau_\sigma$ for all possible choices of $r$ and $s$ we find
 $\braket{{l_\sigma, s} |{r_\lambda, r}}_\tau = 0$ as at least one of the rightmost or leftmost legs of one of the states $_\tau\!\bra{l_\sigma, s}$ or 
 $\ket{r_\lambda, r}_\tau$ is contracted with $\ket{\circ}$.
 The same reasoning yields $\braket{{l_\sigma, s} |{r_\lambda, r}}_\tau = 0$ if $\tau_\lambda = \tau_\sigma$ but $r\neq s$.
 In the remaining case, $\tau_\lambda = \tau_\sigma$ and $r=s$ we have
 $\braket{{l_\sigma, s} |{r_\lambda, r}}_\tau = \braket{l_\sigma | r_\lambda}_{\tau_\lambda}=\delta_{\sigma, \lambda}$ which vanishes if $\sigma \neq  \lambda$ and which is one if $\sigma =  \lambda$ as we can chose $\ket{r_\lambda}_{\tau_\lambda}$ and 
 $_{\tau_\lambda}\!\!\bra{l_\sigma}$ biorthogonal.
 As by Eqs.~\eqref{eq:reduced_eigenvectors1}~and~\eqref{eq:reduced_eigenvectors2} the trivial left and right eigenvectors are 
 orthogonal to all other eigenvectors we conclude that the 
left and right eigenvectors given by Eq.~\eqref{eq:eigenvectors} are biorthogonal and hence
 \begin{align}
     \Proj_{\lambda, \tau}\Proj_{\sigma, \tau} = 
    \sum_{s, r}\ket{r_\lambda, s}\!\bra{l_\lambda, s}_\tau \ket{r_\sigma, r}\! \bra{l_\sigma, l}_\tau = 
     \sum_{s, r}\ket{r_\lambda, s}\!\bra{l_\sigma, r}_\tau\delta_{r,s}\delta_{\sigma, \lambda} = \Proj_{\lambda, \tau} \delta_{\sigma, \lambda}
 \end{align}
 
 This determines the spectral decomposition of the transfer matrix $\T_\tau$ given by Eq.~\eqref{Supp_ex:spectral_decomposition} and allows for determining the
 eigenvalues which contribute to the correlation functions~\eqref{eq:correlation}.
 Inserting the spectral decomposition in 
 Eq.~\eqref{eq:correlations_via_transfer_matrix} we obtain 
  \begin{align}
C_{ab}(t) =  \sum_{\lambda,  \sigma} \lambda^{L-\delta} \sigma^\delta
 c_{ab, \tau}(\lambda, \sigma),
 \label{eq:supp_corrs_decomposition}
 \end{align}
 where the double sum runs over all eigenvalues $\lambda \in \text{spec}(\T_\tau)$ and 
 $\sigma \in \text{spec}(\T_{\tau + 1})$.
 We moreover defined
 \begin{align}
  c_{ab, \tau}(\lambda, \sigma) & := \tr\left(\left[\Proj_{\lambda, \tau} \otimes \mathds{1}_{q^2}\right]\Proj_{\sigma, \tau + 1}\C_{ab, \tau + 1}\right) \\
  & = \sum_{s, r}\tr\left(\left[\ket{r_\lambda, s}\!\bra{l_\lambda, s}_\tau
  \otimes \mathds{1}_{q^2}\right]
  \ket{r_\sigma, r}\!\bra{l_\sigma, r}_{\tau + 1}\C_{ab, \tau + 1}
   \right) \\
  &  =: \sum_{s, r}
   \label{eq:supp_def_cab} c_{ab, \tau}^{s, r}(\lambda, \sigma) 
 \end{align}
 which can be simplified using a diagrammatic representation. To
 this end  we extend the definition~\eqref{eq:eigenvector_diagramm} as
 \begin{align}
 \ket{r_\lambda, s}_{\tau} = 
 \begin{tikzpicture}[baseline=0, scale=0.6]
 \foreach  \x in {0, 1, 2, 4, 5}
 { 
     \draw[thick] (\x, 0.) -- (\x, 0.8);
 }
 \draw[thick, fill=myorange2, rounded corners=1pt] (-0.25, -0.3) rectangle  (0.25 + 5,0.3);
 \Text[x=3.,y=0.55]{$\dots$}
 \Text[x=2.5,y=0]{$\lambda, s$}
 \end{tikzpicture}\quad \text{and} \quad
 _\tau\!\bra{l_\lambda, s} = 
 \begin{tikzpicture}[baseline=-7, scale=0.6]
 \foreach  \x in {0, 1, 2, 4, 5}
 { 
     \draw[thick] (\x, 0.) -- (\x, -0.8);
 }
 \draw[thick, fill=myorange2, rounded corners=1pt] (-0.25, -0.3) rectangle  (0.25 + 5,0.3);
 \Text[x=3.,y=-0.55]{$\dots$}
 \Text[x=2.5,y=0]{$\lambda, s$}
 \end{tikzpicture} \, .
 \end{align}
 Inserting this into $c_{ab, \tau}^{s, r}(\lambda, \sigma)$ for fixed $r$ and $s$ we obtain
 \begin{align}
    c_{ab, \tau}^{s, r}(\lambda, \sigma) & = 
      \begin{tikzpicture}[baseline=-40, scale=0.6]
      \foreach \x in {0, 1, 2, 4, 5, 6}
      {
          \draw[thick] (\x, -5.5 ) arc (180:360: 0.15);
          \draw[thick] (\x, 1. ) arc (360:180: -0.15);
          \draw[thick, color=gray, dashed] (\x + 0.3, -5.5) -- (\x + 0.3, 1.);    
      }
      \foreach \x in {1, 2, 5, 6}
      {
          \Shift{\x - 0.5}{-5.}
      }
     \foreach  \x in {0, 1, 2, 4, 5}
     { 
         \draw[thick] (\x, 0.) -- (\x, 1.);
         \draw[thick] (\x, -1.) -- (\x, -2.5);
     }
     \draw[thick, fill=myorange2, rounded corners=1pt] (-0.25, -0.3) rectangle  (0.25 + 5,0.3);
          \draw[thick, fill=myorange2, rounded corners=1pt] (-0.25, -0.3 - 1.) rectangle  (0.25 + 5,0.3 - 1. );
     \Text[x=3.,y=0.75]{$\dots$}
     \Text[x=2.5,y=0]{$\lambda, s$}
     \Text[x=2.5,y=-1.]{$\lambda, s$}
          \foreach  \x in {0, 1, 2, 4, 5, 6}
     { 
         \draw[thick] (\x, -3.5) -- (\x, -4.5);
     }
 \draw[thick] (6., -2.5) -- (6., 1.);
     \draw[thick, fill=myorange2, rounded corners=1pt] (-0.25, -0.3 - 2.5) rectangle  (0.25 + 6, 0.3 - 2.5);
     \draw[thick, fill=myorange2, rounded corners=1pt] (-0.25, -0.3 - 3.5) rectangle  (0.25 + 6,0.3 - 3.5 );
     \Text[x=3.,y=-1.75]{$\dots$}
     \Text[x=3,y=-2.5- 0.1]{$\sigma, r$}
     \Text[x=3.,y=-3.5- 0.1]{$\sigma, r$}
     \initialOperator{-0.5}{-5.}
     \finalOperator{6.5}{-5.}
     \draw[thick] (2.5, -5.) -- (2.5 - 0.5, -5. - 0.3);
     \draw[thick] (2.5 - 0.5, -5.5) -- (2.5 - 0.5, -5. - 0.3);
     \draw[thick] (3.5 + 0.5, -4.5) -- (3.5 + 0.5, -4.5 - 0.2);
     \draw[thick] (3.5, -5.) -- (3.5 + 0.5, -4.5 - 0.2);
     \Text[x=3.,y=-4.6]{$\dots$}
     \Text[x=6.75,y=-4.5]{$b$}
     \Text[x=-0.75,y=-4.55]{$a$}
     \end{tikzpicture} = 
     \begin{tikzpicture}[baseline=-31, scale=0.6]
     \foreach  \x in {0, 1, 2, 4, 5, 6}
     { 
         \draw[thick] (\x, 0.) -- (\x, -1.);
     }
     \draw[thick, fill=myorange2, rounded corners=1pt] (-0.25, -0.3) rectangle  (0.25 + 5,0.3);
     \draw[thick, fill=myorange2, rounded corners=1pt] (-0.25, -0.3 - 1.) rectangle  (0.25 + 6,0.3 - 1. );
     \localOperator{6.}{0.}
     \Text[x=6.4,y=0.]{$b$}
     \Text[x=3.,y=-0.5]{$\dots$}
     \Text[x=2.5,y=0]{$\lambda, s$}
     \Text[x=3,y=-1.- 0.1]{$\sigma, r$}
          \foreach  \x in {0, 1, 2, 4, 5, 6}
     { 
         \draw[thick] (\x, -2) -- (\x, -3.);
     }
     \draw[thick, fill=myorange2, rounded corners=1pt] (-0.25 + 1., -0.3 - 3) rectangle  (0.25 + 6.,0.3 - 3.);
     \draw[thick, fill=myorange2, rounded corners=1pt] (-0.25, -0.3 - 2.) rectangle  (0.25 + 6,0.3 - 2. );
     \localOperator{0.}{-3.}
     \Text[x=-0.4,y=-3.]{$a$}
     \Text[x=3.,y=-2.5]{$\dots$}
     \Text[x=3.5,y=-3.]{$\lambda, s$}
     \Text[x=3,y=-2.- 0.1]{$\sigma, r$}
     \end{tikzpicture} \\
     & = \left( _\tau\!\bra{l_\lambda, s}\otimes \, _1\!\!\bra{b}\right) \ket{r_\sigma, r}\!
     \bra{l_\sigma, r}_{\tau + 1}\left( \ket{a}_1 \otimes \ket{r_\lambda, s}_{\tau}\right).
     \label{eq:contributions_from_eigenvectors}
 \end{align}

We aim for showing that the latter expression is nonzero only if $\tau_\lambda = \tau$ and
$\tau_\sigma = \tau + 1$ and hence the double sum over $r$ and $s$ in Eq.~\eqref{eq:supp_def_cab} reduces to a single term.
On the one hand the condition on $\tau_\sigma$ is due to the local operators $a$ and $b$ being traceless.
If the left (right) eigenvector corresponding to $\sigma$ has a factor $\ket{\circ}_1$ on the first (last) site this produces as factor $\braket{\circ | a}_1 = 0$ 
($\braket{b |\circ}_1 = 0$).
On the other hand the condition on $\tau_\lambda$ is a consequence of the properties~\eqref{eq:reduced_eigenvectors1}~and~\eqref{eq:reduced_eigenvectors2}.
If the left (right) eigenvector corresponding to $\lambda$ has a factor $\ket{\circ}$ on the first (last) site contraction along the corresponding wire maps $\ket{r_\sigma}_{\tau + 1}$ ($ _{\tau + 1}\!\!\bra{l_\sigma}$) to $\ket{r_\sigma, -}_{\tau} = 0$
($ _\tau\!\bra{l_\sigma, +} = 0$).
This argument might be formalized as follows:
First, assume that $\tau_\sigma < \tau + 1$ and thus 
$r \in \{0, 1, \ldots, \tau + 1 - \tau_\sigma\}$.
If $r>0$ we might write 
$_{\tau + 1}\!\!\bra{l_\sigma, r}= \, _1\!\!\bra{\circ}\otimes\,  _\tau\!\bra{l_\sigma, r-1}$ and hence the second factor in
Eq.~\eqref{eq:contributions_from_eigenvectors} becomes 
\begin{align}
   _{\tau+1}\!\!\bra{l_\sigma, r}\left( \ket{a}_1 \otimes \ket{r_\lambda, s}_{\tau}\right) =
   \braket{\circ | a }_1 \braket{l_\sigma, r-1| r_\lambda, s}_{\tau} = 0
\end{align}
as we have chosen the local operator $a$ to be traceless and therefore 
Hilbert-Schmidt orthogonal to the identity which in verctorized form reads $\braket{\circ | a }_1 = 0$.
If on the other hand $r=0$ we might repeat the argument and write 
$\ket{r_\sigma, r}_{\tau + 1}= \ket{r_\sigma, r}_\tau \otimes \ket{\circ}_1$ for which the first factor becomes 
\begin{align}
\left( _\tau\!\bra{l_\lambda, s} \otimes \, _1\!\!\bra{b}\right) \ket{r_\sigma, r}_{\tau + 1} = \braket{l_\lambda, s|r_\sigma, r}_\tau \braket{b|\circ}_1 = 0
\end{align}
 as $b$ is traceless.
 This proofs that $c_{ab, \tau}(\lambda, \sigma) \neq 0$ only if $\tau_\sigma = \tau + 1$ and hence $\ket{r_\sigma, r}_{\tau + 1}$ ($ _{\tau+1}\!\!\bra{l_\sigma, r}$) can be replaced by $\ket{r_\sigma}_{\tau + 1}$ ($ _\tau\!\bra{l_\sigma}$).
 We continue by demonstrating that for $c_{ab, \tau}^{s, r}(\lambda, \sigma) \neq 0$ and hence 
 $c_{ab, \tau}(\lambda, \sigma) \neq 0$
 also $\tau_\lambda = \tau$ has to hold by similar reasoning.
 Assume, that this is not the case and therefore $s \in \{0, 1, \ldots \tau - \tau_\lambda\}$.
 If $s>0$ we write $_\tau\!\bra{l_\lambda, s} = \,
_1\!\!\bra{\circ} \otimes \, _{\tau - 1}\!\! \bra{l_\lambda, s-1}$ and obtain
\begin{align}
    \left( _\tau\!\bra{l_\lambda, s}\otimes\, _1\!\!\bra{b}\right)
    \ket{r_\sigma}_{\tau + 1} = 
        \left( _1\!\!\bra{\circ} \otimes \, _{\tau - 1}\!\! \bra{l_\lambda, s-1} \otimes \,
        _1 \!\!\bra{b}\right) \ket{r_\sigma}_{\tau + 1} = 
         \left( _{\tau - 1}\!\!\bra{l_\lambda, s-1} \otimes\,_1\!\!\bra{b}\right) \ket{r_\sigma, -}_{\tau} = 0,     
\end{align}
where we used Eq.~\eqref{eq:reduced_eigenvectors1}.
Similarly, for $s=0$ we write 
 $\ket{r_\lambda, s}_\tau =
 \ket{r_\lambda, s}_{\tau -1} \otimes \ket{\circ}_1$ and obtain
\begin{align}
    _{\tau + 1}\!\!\bra{l_\sigma}\left( \ket{a}_1 \otimes \ket{r_\lambda, s}_{\tau}\right) = 
    \, _{\tau + 1}\!\!\bra{l_\sigma}\left( \ket{a}_1 \otimes \ket{r_\lambda, s}_{\tau - 1}  \otimes \ket{\circ}_1\right) =
    \, _\tau\!\bra{l_\sigma, +}\left( \ket{a}_1 \otimes \ket{r_\lambda, s}_{\tau - 1}\right) = 0.
\end{align}
Therefore we conclude that $c_{ab, \tau}(\lambda, \sigma) \neq 0$ only if $\tau_\lambda = \tau$ and hence $\ket{r_\lambda, s}_{\tau}$ ($ _\tau\!\bra{l_\lambda, s}$) can be replaced by $\ket{r_\lambda}_{\tau}$ ($ _\tau\!\bra{l_\lambda}$) yielding 
\begin{align}
    c_{ab, \tau}(\lambda, \sigma) = 
    \delta_{\tau_\lambda,\tau}\delta_{\tau_\sigma,\tau + 1}
   \left( \, _\tau\!\bra{l_\lambda} \otimes\,_1\!\!\bra{b}\right) 
   \ket{r_\sigma}\!
    \bra{l_\sigma}_{\tau + 1}\left( \ket{a}_1 \otimes \ket{r_\lambda}_{\tau}\right).
\end{align}
Inserting the latter expression back into Eq.~\eqref{eq:supp_corrs_decomposition} we finally obtain Eq.~\eqref{eq:correlations_from_spectral_decomposition}.

\section{Asymptotics of correlations for generic impurities}
\label{S-VI}

In this section we derive the asymptotic expression~\eqref{eq:asymptotic_correaltion_generic_1}
for correlation functions in the case of generic impurities.
Our numerical experiments show
that the transfer matrix $\T_\tau$ in general fails to be diagonalizable if $\tau > 1$ and the nontrivial
eigenvalues lead to nontrivial Jordan blocks.
Hence we start by describing the Jordan decomposition of $\T_\tau$ in the following.
Using the notation from the T-dual case, Eq.~\eqref{eq:eigenvectors}, only $\ket{r_\lambda, \tau-\tau_\lambda}$ ($\bra{l_\lambda, 0}$) is a right (left) eigenvector with eigenvalue $\lambda$.
The corresponding Jordan block, again assuming no accidental degeneracies, has dimension $d_\lambda = \tau-\tau_\lambda + 1$ and corresponds to the $\T_\tau$ invariant subspace
$\ker\!\left(\N_{\lambda, \tau}^{d_\lambda}\right)$,
where $\N_{\lambda, \tau} = \T_\tau - \lambda \mathds{1}_{q^{2\tau}} $.
Let us denote the projection onto this subspace by $\Proj_{\lambda, \tau}$, which does not coincide with Eq.~\eqref{eq:proj_dual_unitary}.
Note that $\N_{\lambda, \tau}^{d_\lambda} \Proj_{\lambda, \tau} = 0$ and in particular for the trivial eigenvalue $\N_{1, \tau}\Proj_{1, \tau} = 0$.
The projections $\Proj_{\lambda, \tau}$ again form a resolution of identity and we write 
\begin{align}
\T_{\tau} = \sum_{\lambda \in \text{spec}(\T_\tau)} \left(\lambda \mathds{1}_{q^{2\tau}} + \N_{\lambda, \tau}\right)\Proj_{\lambda, \tau} = \ket{\circ}\!\bra{\circ}^{\otimes \tau} +
\sum_\lambda \left(\lambda \mathds{1}_{q^{2\tau}} + \N_{\lambda, \tau}\right)\Proj_{\lambda, \tau},
\end{align}
where the sums on the right hand side and in what follows runs only over the nontrivial 
eigenvalues $\lambda \in 
\text{spec}(\T_\tau)$.
As we are interested in the scaling of correlations with system size $L$ let us assume that
$L-\delta \gg  \tau$ and hence $L -\delta > d_\lambda$ for all $\lambda \in \text{spec}(\T_\tau)$
for fixed $\delta$.
Upon inserting the Jordan decomposition into Eq.~\eqref{eq:correlations_via_transfer_matrix} we obtain
\begin{align}
C_{ab}(t)& = \tr\left( \left[\ket{\circ}\!\bra{\circ}^{\otimes \tau} \otimes \mathds{1}_{q^2} \right] \T_{\tau + 1}^\delta\C_{ab, \tau+1}\right) + \,
\sum_\lambda \tr\left(\left[\left(\lambda \mathds{1}_{q^{2\tau}} + \N_{\lambda, \tau}\right)^{L-\delta}\Proj_{\lambda, \tau} \otimes \mathds{1}_{q^2} \right]
\T_{\tau + 1}^\delta\C_{ab, \tau+1}\right) \\
& = \bra{\circ}^{\otimes \tau} \otimes \bra{b} \T_{\tau+1}^\delta
\ket{a} \otimes \ket{\circ}^{\otimes \tau}  + \,
\sum_\lambda \lambda^{L-\delta - d_\lambda} 
\sum_{n=0}^{d_\lambda}\binom{L-\delta}{n}
\lambda^{d_\lambda - n} \tr\left(\left[ 
\N_{\lambda, \tau}^n \Proj_{\lambda, \tau}
\otimes \mathds{1}_{q^2}
\right] \T_{\tau + 1}^\delta \C_{ab, \tau+1} \right).
\label{eq:correlations_generic_spectral_decomposition2}
\end{align}
Assuming a spectral gap, as suggested by our numerical experiments, the second term in the last line
vanishes as $L^\tau\lambda_0^L$ as $L$ tends to infinity, with the factor $L^\tau$ bounding the asymptotic growth of the binomial coefficients $\binom{L-\delta}{n}$.
Hence asymptotically the second term is exponentially suppressed as $\lambda_0^L$ and the first term which coincides with Eq.~\eqref{eq:asymptotic_correaltion_generic_1} governs the correlation function.
Let us further explore the $\delta$-dependence of the second term.
To this end we assume $\delta \gg \tau+1$, i.e., $\delta$ larger than the dimension of the largest
Jordan block.
By inserting the Jordan decomposition of $\T_{\tau + 1}$ into the second term of the right hand side in Eq.~\eqref{eq:correlations_generic_spectral_decomposition2}
we obtain for the trace 
\begin{align}
 \tr\left(\left[ 
\N_{\lambda, \tau}^n \Proj_{\lambda, \tau}
\otimes \mathds{1}_{q^2}
\right] \T_{\tau + 1}^\delta \C_{ab, \tau+1} \right)
& = 
\sum_{\sigma}\sigma^{\delta - d_\sigma} 
\sum_{m=0}^{d_\sigma}\binom{\delta}{m}\sigma^{d_\sigma - m}
\tr\left(\left[ 
\N_{\lambda, \tau}^n \Proj_{\lambda, \tau}
\otimes \mathds{1}_{q^2}
\right] 
\N_{\sigma, \tau+1}^m \Proj_{\sigma, \tau+1} \C_{ab, \tau+1} 
\right).
\end{align}
Here the sum over $\sigma$ runs again over nontrivial eigenvalues only as
$\ket{\circ}\!\bra{\circ}^{\otimes \tau + 1}\C_{ab, \tau + 1}=0$.
Consequently, the above expression 
is dominated by the leading nontrivial eigenvalues $\sigma_0$ of $\T_{\tau + 1}$ 
for large $\delta$ and hence
the second term in Eq.~\eqref{eq:correlations_generic_spectral_decomposition2} is suppressed
at least as $|\sigma_0|^L$ for all $\delta$ since $|\lambda_0|\leq |\sigma_0|$. Again the binomial coefficient gives only a polynomial correction proportional to $\delta^\tau$.
Hence the first term dominates the large $L$ asymptotics of correlation functions.
Inserting the Jordan decomposition of $\T_{\tau + 1}$ there and noting that 
$\ket{\circ}^{\otimes \tau + 1}$ is orthogonal to $\ket{\circ}^{\otimes \tau}\otimes \ket{a}$
yields
\begin{align}
 \bra{\circ}^{\otimes \tau} \otimes \bra{b} \T_{\tau+1}^\delta
\ket{a} \otimes \ket{\circ}^{\otimes \tau} = 
\sum_{\sigma}\sigma^{\delta - d_\sigma} 
\sum_{m=0}^{d_\sigma}\binom{\delta}{m}
\sigma^{d_\sigma - m}
\bra{\circ}^{\otimes \tau} \otimes \bra{b}
\N_{\sigma, \tau+1}^m \Proj_{\sigma, \tau+1} \ket{a} \otimes \ket{\circ}^{\otimes \tau}.
\end{align}
By the same reasoning as above this is dominated by the leading nontrivial eigenvalue
$\sigma_0$ of $\T_{\tau+1}$ for sufficiently large $\delta$.
Hence one has the asymptotics 
$C_{ab}(t=\tau L + \delta) \sim \sigma_0(\tau)^\delta$ if both $\delta$ and $L-\delta$ are large.
Here, we indicate the $\tau$ dependence of $\sigma_0$ in the notation and for simplicity assume a unique real leading eigenvalue.
Note, that our numerical experiments on the spectral properties of the transfer matrices
imply $|\sigma_0| \approx 1/q$.
As indicated in the main text at times $t \approx \tau L$ non-universal corrections to the asymptotic scaling need to be taken into account.

Using the diagrammatic representation~\eqref{eq:asymptotic_correaltion_generic_2} of the leading contributions to the correlation function suggests that the correlations should exponentially decay with $\tau$.
This can be seen by evaluating the diagram in terms of column transfer matrices instead of the (row) transfer matrices $\T_{\tau + 1}$. 
Both types of transfer matrices share similar spectral properties and hence the diagram gives a
contribution scaling with the largest nontrivial eigenvalue of the column transfer
matrix $\chi_0$ as $\chi_0^\tau$.
The arguments outlined above, however, do not guarantee that the second term in 
Eq.~\eqref{eq:correlations_generic_spectral_decomposition2} gives rise only to subleading terms when $L$ is kept fixed and $\tau \to \infty$.

\end{widetext}

%

\end{document}